\documentclass[journal,draftclsnofoot,onecolumn,12pt,twoside]{IEEEtran}
\IEEEoverridecommandlockouts
\usepackage{amsmath}
\usepackage{amsfonts}
\usepackage{amssymb}
\usepackage{graphicx}
\usepackage{epstopdf}
\usepackage{caption}
\usepackage{subcaption}
\usepackage{cite}
\usepackage{algorithmicx}
\usepackage[ruled]{algorithm}
\usepackage{algpseudocode}
\usepackage{blindtext}
\usepackage{enumitem}
\usepackage{pdfpages}
\usepackage{bbm}
\usepackage{bibentry}
\usepackage{amsthm}

\usepackage{changes}
\definechangesauthor[color=blue,name={Mingyue Ji}]{MJ}

\ifodd 1

\else

\fi

\ifodd 1

\else

\fi

\renewcommand{\baselinestretch}{1.4}

\begin{document}
%
\title{Throughput-Outage Scaling Behaviors for Wireless Single-Hop D2D Caching Networks with Physical Model -- Analysis and Derivations}

\author{Ming-Chun Lee,~\IEEEmembership{Member,~IEEE}, Andreas F. Molisch,~\IEEEmembership{Fellow,~IEEE}, and Mingyue Ji,~\IEEEmembership{Member,~IEEE}
\thanks{M.-C. Lee is with Institute of Communications, National Yang Ming Chiao Tung University and National Chiao Tung University, Hsinchu 30010, Taiwan. (email: mingchunlee@nctu.edu.tw)}
\thanks {A. F. Molisch is with Department of Electrical and Computer Engineering, University of Southern California, Los Angeles, CA 90089, USA (email: molisch@usc.edu).}
\thanks{M. Ji is with Department of Electrical and Computer Engineering, University of Utah, Salt Lake City, UT 84112, USA (email: mingyue.ji@utah.edu).}
\thanks{Part of this work will be presented in the 2021 IEEE International Conference on Communications \cite{lee2021throughput}. A condensed version of this paper has been submitted to the IEEE Transactions on Communications.}
}


%


\newcommand{\eqdef}{=\vcentcolon}

\maketitle
\begin{abstract}
Throughput-Outage scaling laws for single-hop cache-aided device-to-device (D2D) communications have been extensively investigated under the assumption of the protocol model. However, the corresponding performance under physical models has not been explored; in particular it remains unclear whether link-level power control and scheduling can improve the asymptotic performance. This paper thus investigates the throughput-outage scaling laws of cache-aided single-hop D2D networks considering a general physical channel model. By considering the networks with and without the equal-throughput assumption, we analyze the corresponding outer bounds and provide the achievable performance analysis. Results show that when the equal-throughput assumption is considered, using link-level power control and scheduling cannot improve the scaling laws. On the other hand, when the equal-throughput assumption is not considered, we show that the proposed double time-slot framework with appropriate link-level power control and scheduling can significantly improve the throughput-outage scaling laws, where the fundamental concept is to first distinguish links according to their communication distances, and then enhance the throughput for links with small communication distances.
\end{abstract}


%
\IEEEpeerreviewmaketitle

\section{Introduction}

In the past few years, the demand of video services has increased rapidly for mobile devices \cite{Cisco_2017}, and thus significant efforts have been made to deal with such challenge \cite{wang2017survey,zhang20196g}. Although the improvement from conventional approaches \cite{Andrews:5G}, e.g., use of additional spectrum, including massive antenna systems, and adopting network densifications, can partially resolve the challenge, the improvement might still be inadequate and inefficient \cite{Gol:femtocaching}. In this context, caching at the wireless edge was introduced and investigated to further improve the network performance \cite{Gol:femtocaching}.\footnote{Note that caching at the wireless edge is not a competing technology for the conventional approaches. On the contrary, it can complement the conventional approaches, and thus further improve the network performance.} 

The fundamental concept of caching at the wireless edge is to convert memory to bandwidth by pre-fetching and caching the popular video content at the network edge nodes so that the video content can be rapidly delivered to users with low cost when demanded \cite{Gol:femtocaching}. Thus, this along with the concentration of the popularity can bring huge benefits to the network. Recent investigations have revealed that by introducing the caching technologies at the wireless edge, the network can be improved by orders of magnitude in practical simulations \cite{Ji:Dcache,li2018survey} and alter the fundamental scaling in theoretical analysis \cite{Maddah-Ali:CCache,Ji:Th_Out_toff,lee2019throughput,lee2020optimal}.

Due to the significance of having caching at the wireless edge, the approach has been investigated in different scenarios \cite{li2018survey,ahmed2019video,mehrabi2019device,prerna2020device}. These include caching in BSs \cite{Gol:femtocaching}, caching on devices \cite{Ji:Dcache}, caching in heterogeneous networks \cite{li2017collaborative}, caching in vehicles \cite{hu2018mobility}, etc. As caching on devices along with high performance device-to-device (D2D) communications can bring huge benefits to the network without installing new infrastructure, cache-aided wireless D2D networks have been widely discussed in the literature \cite{ahmed2019video,mehrabi2019device,prerna2020device}. While many papers investigate the design and implementation aspects, another set of investigations focuses on characterizing the asymptotic behavior as the number of users $N$ goes to infinity. Such investigations are commonly referred to as scaling law analysis, where the analysis results can be used to understand the fundamental limits and benefits of the network \cite{Ji:Th_Out_toff,lee2019throughput,lee2020optimal}. This paper aims to contribute to this range of investigations for cache-aided wireless D2D networks.

\subsection{Related Literature}

The scaling laws for wireless D2D/ad-hoc networks without caching have been studied for many years since the seminal work in \cite{gupta2000capacity}.\footnote{Scaling law order notations: given two functions $f$ and $g$, we say that: (1) $f(n)=\mathcal{O}(g(n))$ if there exists a constant $c$ and integer $N$ such that $f(n)\leq cg(n)$ for $n>N$. (2) $f(n)=o(g(n))$ if $\displaystyle{\lim_{n\to\infty}}\frac{f(n)}{g(n)}=0$. (3) $f(n)=\Omega(g(n))$ if $g(n)=\mathcal{O}(f(n))$. (4) $f(n)=\omega(g(n))$ if $g(n)=o(f(n))$. (5) $f(n)=\Theta(g(n))$ if $f(n)=\mathcal{O}(g(n))$ and $g(n)=\mathcal{O}(f(n))$.} Known as one of most representative papers for scaling law analysis, \cite{gupta2000capacity} studied the transport capacity of the network under protocol and physical models and characterized the throughput scaling laws, where the derived achievable throughput and upper bound in the case of adopting the multi-hop D2D communication were $\Theta\left(\frac{1}{\sqrt{N\log(N)}}\right)$ and $\Theta\left(\frac{1}{\sqrt{N}}\right)$, respectively. Refs \cite{agarwal2004capacity} and \cite{xue2006scaling} extended the analysis and showed that the upper bound is $\Theta\left(\frac{1}{\sqrt{N}}\right)$ even under more general conditions. When users are distributed according to a Poisson point process, \cite{franceschetti2007closing} showed that the optimal achievable throughput for wireless D2D networks is $\Theta\left(\frac{1}{\sqrt{N}}\right)$, matching the upper bound suggested in  \cite{agarwal2004capacity} and \cite{xue2006scaling}. In addition to the above mentioned papers, scaling law analysis has been conducted for networks with more complicated settings and/or channel models. For example, analysis considering fading was dicussed in \cite{xue2006scaling}; analysis considering a specific user mobility model was provided in \cite{el2006optimal_I}; analysis considering multicasting benefit was presented in \cite{shakkottai2010multicast}; and analysis involving a special distributed multiple-input multiple-output (MIMO) structure, known as hierarchical cooperation, was introduced in \cite{ozgur2007hierarchical}.

Although cache-aided D2D networks have been investigated for years by the computer science community \cite{hara2001effective,cohen2002replication,hara2006data,zhao2010cooperative}, their fundamental scaling behaviors were not the focus until the early 2010s. Known as one of the earliest scaling law analysis for cache-aided D2D networks, \cite{Golrezaei:Cach_scale} characterized the scaling law of the maximum expected throughput in cache-aided single-hop D2D networks under the protocol model. Since focusing on the maximum expected throughput characterization ignores the outage probability analysis, \cite{ji2013fundamental} and \cite{Ji:Th_Out_toff} remedied this aspect and provided the throughput-outage scaling law analysis again for cache-aided single-hop D2D with protocol model. The results showed that when the outage probability is negligibly small and the popularity distribution is a heavy-tailed Zipf distribution, the throughput scales with $\Theta\left(\frac{S}{M}\right)$, where $M$ is the number of contents in the library and $S$ is the cache space of a device. By considering the more practical MZipf popularity distribution, the results of \cite{Ji:Th_Out_toff} were then generalized in \cite{lee2019throughput}.

Cache-aided multihop D2D scaling laws were firstly analyzed in \cite{gitzenis2012asymptotic}, where the average traffic per node was characterized with users located on a regular grid and with a fairly simplified channel model. Then, the throughput-outage scaling law analysis was provided in \cite{jeon2017wireless} under the protocol model. The results showed that when the outage probability is vanishing, the achievable throughput per user is $\Theta\left(\sqrt{\frac{S}{M\log(N)}}\right)$ for heavy-tailed Zipf popularity distributions, while the upper bound is $\Theta\left(\sqrt{\frac{S\log(N)}{M}}\right)$. This upper bound was improved to $\Theta\left(\sqrt{\frac{S}{M}}\right)$ in \cite{qiu2019popularity}, where the more practical physical model was used and a fully centralized caching policy was considered. Considering the pros and cons in \cite{gitzenis2012asymptotic,jeon2017wireless,qiu2019popularity}, \cite{lee2020optimal} studied the cache-aided multihop D2D scaling laws adopting the Poisson point process (PPP) for the user distribution, physical model for transmissions, decentralized policy for caching, and MZipf distribution for popularity. Results in \cite{lee2020optimal} demonstrated that when the outage probability is negligibly small, the optimal throughput per user is $\Theta\left(\sqrt{\frac{S}{M}}\right)$ for heavy-tailed MZipf distribution and is $\Theta\left(\sqrt{\frac{S}{q}}\right)$ for light-tailed MZipf distribution, where $q$ is the plateau factor of the MZipf distribution.

The above literature considered conventional single-hop and multihop D2D for communications with users more or less uniformly distributed within the network, though there exist papers also considering more complicated settings and communication schemes. For example, in \cite{guo2017achievable}, a scaling law analysis was conducted for the cache-aided hierarchical cooperation approach. Besides, the scaling law in cache-aided D2D networks considering nonuniform user distribution was investigated in \cite{ren2021scaling}. Moreover, when involving coding and multicasting schemes, coded cache-aided D2D was proposed and analyzed in \cite{ji2017fundamental,Ji:Fund_D2D,naderializadeh2017fundamental,yapar2019optimality,zhang2021cache}.

\subsection{Contributions}

In this paper, we focus on the scaling law analysis for cache-aided single-hop D2D networks. By the above literature review, we observe that the scaling law investigations for single-hop cache-aided D2D networks, i.e., \cite{Golrezaei:Cach_scale,Ji:Th_Out_toff,lee2019throughput}, were conducted mostly with protocol model. However, the protocol model might be oversimplified, as it cannot incorporate the influence of link-level power control and scheduling into the analysis. A more realistic model to use is the physical model \cite{agarwal2004capacity}, in which the influence of link-level power control and scheduling can be accommodated. Although the suitable scheduling and power control algorithms have been investigated for finite-size networks and performance has been investigated by simulations \cite{Zhang:D2D_Schedule,chen2017optimal,lee2018cachingTWC,choi2020joint}, to the best of our knowledge, the scaling behaviors for cache-aided single-hop D2D networks with physical model have not been explored, and it is unclear whether and how the link-level power control and scheduling can further improve the scaling laws as compared with those derived under protocol model. This paper thus aims to contribute to this aspect.

Specifically, this paper considers a single-hop cache-aided D2D network with MZipf popularity distribution and with users to be uniformly distributed in the network. We conduct the throughput-outage scaling law analysis with the generalized physical model.\footnote{Although the scaling laws of the bounded physical model could be more realistic, as indicated in \cite{agarwal2004capacity}, they can be viewed as special cases of the scaling laws derived considering the generalized physical channel. Therefore, we in this paper focus on the generalized physical model for brevity. This assumption will discussed in detail in Sec. \ref{Sec:Net_Setup}.A.} 
We consider two fundamental scenarios for the networks, where scenario 1 assumes that the link-level throughput realizations are equal for all users in the network and scenario 2 relaxes the assumption such that different users can have different link-level throughput realizations, leading to certain unfairness. It should be noted that since the unfairness of the second scenario is only on the realization level, users in either scenario on average have the same throughput, namely, users are treated {\em statistically fair} in both scenarios. 

We consider the regime that the outage probability is negligibly small or converging to zero for the analysis, and conduct both the achievable and outer bound analysis assuming $M,N\to\infty$. Specifically, when the MZipf distribution is heavy-tailed, i.e., $\gamma<1$ and $q\to\infty$, where $\gamma$ is the Zipf factor of the distribution, we show that the upper bound of the throughput per user for scenario 1 is $\Theta\left(\frac{S}{M}\right)$ when the outage probability is negligibly small, which is identical to the throughput-outage scaling laws assuming the protocol model \cite{lee2019throughput}. This indicates that the performance of single-hop cache-aided D2D networks cannot be asymptotically improved by link-level power control and scheduling in scenario 1. Note that the assumption that $q\to\infty$ is to avoid the situation that the MZipf distribution degenerates to a Zipf distribution asymptotically. In contrast, we demonstrate that in scenario 2, by using the proposed throughput-enhancing approach, the throughput per user upper bound can be enhanced to $\Theta\left(\left(\frac{S}{M}\right)^{\frac{1-\gamma}{2-\gamma}}\right)$ while the outage probability retains negligibly small. The fundamental concept of the proposed throughput-enhancing approach is to let transmitter (TX)-receiver (RX) pairs with small communication distances to communicate with a very high speed under appropriate link-level power control and scheduling. In addition to the outer bound analysis, the achievable schemes for both scenarios 1 and 2 adopting $\gamma<1$ are proposed, and the analysis shows that the proposed schemes can achieve their corresponding outer bounds.

We conduct the analysis also for the light-tailed MZipf distribution, where $\gamma>1$ and $q\to\infty$ is considered. We show in this case that the user throughput upper bound is $\Theta\left(\frac{S}{q}\right)$ with negligibly small outage probability for scenario 1, again indicating that using link-level power control and scheduling cannot improve the throughput-outage scaling law in scenario 1. On the other hand, we demonstrate that the throughput upper bound can be enhanced to $\Theta\left(\sqrt{\frac{S}{q}}\right)$ with negligibly small outage probability by the proposed throughput-enhancing approach for scenario 2. The schemes that can achieve these outer bounds are proposed and analyzed, respectively. Finally, we analyze the throughput-outage scaling law considering the Zipf distribution under the conditions that $\gamma>1$ and that the maximum instantaneous power can go to infinity with average user power remaining constant. This is to see whether allowing the maximum instantaneous power going to infinity can allow us to break the $\Theta(1)$ throughput limitation in the physical model when the instantaneous power is upper bounded by some constant. However, the result shows that this is not possible even though we allow the instantaneous power going to infinity with the average user power still being some constant.

\subsection{Paper Organization}

The remainder of this paper is organized as follows. Sec. II discusses the models, assumptions, scenarios, and definitions of the throughput and outage adopted in this paper. Sec. III provides the throughput-outage scaling law analysis considering the MZipf distribution with $\gamma<1$. The scaling law analysis considering the MZipf distribution with $\gamma>1$ is provided in Sec. IV. The analysis for the Zipf distribution with $\gamma>1$ and infinite maximum instantaneous power is presented in Sec. V. Conclusions and some discussions of this paper are provided in Sec. VI. The proofs are relegated to appendices at the end of this paper.

\section{Network Model}

\label{Sec:Net_Setup}

We consider a random dense network where users are placed according to a binomial point process (BPP) within a unit square-shaped area $[0,1]\times [0,1]$. Accordingly, we assume that the number of users in the network is $N$, and users are distributed uniformly at random within the network. We assume each device in the network can cache $S$ files and each file has equal size. We consider a library consisting of $M$ files. We assume that users request the files from the library independently according to a request distribution modeled by the MZipf distribution \cite{lee2019throughput}:
\begin{equation}
P_r(f;\gamma,q)=\frac{(f+q)^{-\gamma}}{\sum_{m=1}^M (m+q)^{-\gamma}}=\frac{(f+q)^{-\gamma}}{H(1,M,\gamma,q)},
\end{equation}
where $\gamma$ is the Zipf factor; $q$ is the plateau factor of the distribution; and $H(a,b,\gamma,q):=\sum_{f=a}^b (f+q)^{-\gamma}$. Note that the MZipf distribution degenerates to a Zipf distribution when $q=0$. To simplify the notation, we will in the remainder of this paper use $P_r(f)$ instead of $P_r(f;\gamma,q)$ as the short-handed expression. We consider a decentralized random caching policy for all users \cite{Blaszczyszyn:fcache}, in which users cache files independently according to the same caching policy. Denoting $P_c(f)$ as the probability that a user caches file $f$, the caching policy is fully described by $P_c(1),P_c(2),...,P_c(M)$, where $0\leq P_c(f)\leq 1,\forall f$; thus users cache files according to the caching policy $\lbrace P_c(f) \rbrace_{f=1}^M$. We consider $\sum_{f=1}^M P_c(f)=S$. Then, each user can cache exactly $S$ different files according to the caching mechanism provided in \cite{Blaszczyszyn:fcache}. In this paper, we assume that $S$ and $\gamma$ are some constants.

We consider the asymptotic analysis in this paper, in which we assume that $N\to\infty$ and $M\to\infty$. We will restrict to $M=o(N)$ and $q=\mathcal{O}(M)$ when $\gamma<1$; $M=o(N)$ and $q=o(M)$ when $\gamma>1$. The main reason for restricting to $M=o(N)$ when $\gamma<1$ is to let users of the network have sufficient ability to cache the whole library. Similarly, the assumption that $q=o(M)$ and $M=o(N)$ when $\gamma>1$ can give the users of the network a sufficient ability to cache the most popular $q$ files (orderwise); otherwise the outage probability would go to $1$.

The plateau factor $q$ can either go to infinity or remain constant. When $q$ goes to infinity, it is sufficient to consider $q=\mathcal{O}(M)$. This is because the MZipf distribution would behave like a uniform distribution asymptotically as $q=\omega(M)$ and such case is
less interesting because the concentration property of files in this case is not captured and because this is equivalent to letting $\gamma$ very close to 0. Consequently, we assume $q=\mathcal{O}(M)$ when $\gamma<1$. In addition, when $\gamma>1$, it is more interesting to consider the case that $q=o(M)$ because it gives a clear distinction between the heavy-tailed case ($\gamma<1$) and the light-tailed case ($\gamma>1$), where the mathematical definition of a heavy-tailed popularity distribution can be found in Definition 3 of \cite{jeon2017wireless}. Furthermore, in practical terms, we see from the measurement results \cite{lee2019throughput} that $q$ is much smaller than $M$ when $\gamma>1$, which supports the consideration of $q=o(M)$. When $q$ is a constant, i.e., $q=\Theta(1)$, the request distribution generally behaves like a Zipf distribution as $M\to\infty$. Thus, the results for $q=\Theta(1)$ can be representative for the analysis that uses the Zipf distribution for the request distribution. We will consider $q\to\infty$ in Secs. \ref{Sec:Out_Bound_Gen} and \ref{Sec:Out_Bound_Gen_gg1} and consider $q=\Theta(1)$ only in Sec. \ref{Sec:Zipf_gg1}. 

We consider single-hop D2D communications for file delivery. We assume users can obtain their desired files through only single-hop D2D communications and assume users always have requests to satisfy. Note that we do not eliminate the possibility that a user can find the desired file from its own cache, and such case can be accommodated by letting the distance between the TX and RX be much smaller than the general D2D communication distance. However, we note that since $S$ is some constant, the probability that a user can find the desired file from its own cache goes to zero as $q$ and $M$ go to infinity. Furthermore, as we would assume the link-rate for file delivery is upper bounded by the power of the TX, the self-caching gain is indeed not significant in terms of asymptotic performance. Similar to \cite{Ji:Th_Out_toff}, we assume that different users making the requests on the same file would request different segments of the file. This avoids the gain from naive multicasting.

We define an outage as an occurrence where a user cannot obtain its desired file from the D2D network. Suppose we are given a realization of the placement of the user locations $\mathsf{P}$ according to the binomial point process. In addition, we are given a realization of file requests $\mathsf{F}$ and a realization of file placement $\mathsf{G}$ of users according to the popularity distribution $P_r(\cdot)$ and caching policy $P_c(\cdot)$, respectively. We can define $T_u$ as the throughput of user $u\in\mathcal{U}$ under a feasible single-hop file delivery scheme. Therefore, $T_u$ is defined as:
\begin{equation}
T_u=\frac{1}{T}\sum_{t=1}^T C_{u}(t)A_u(t),
\end{equation}
where $T$ is the number of time-slots for the transmission, $C_{u}(t)$ is the link rate for user $u$ in time-slot $t$, and $A_u(t)$ is the link activation indicator of user $u$ at time-slot $t$, where $A_u(t)=1$ if the link of user $u$ is scheduled at time-slot $t$; otherwise $A_u(t)=0$. We then define the average throughput of user $u$ as $\overline{T}_u=\mathbb{E}_{\mathsf{P},\mathsf{F},\mathsf{G}}[T_u]$, where the expectation is taken over the placement of user locations $\mathsf{P}$, file requests $\mathsf{F}$ of users, the file placement of users $\mathsf{G}$, and the file delivery scheme. Finally, we define the average throughput of a user in the network as
\begin{equation}\label{eq:throu_def}
T=\displaystyle{\min_{u\in\mathcal{U}}\overline{T}_{u}}.
\end{equation}

When the number of users in the network is $N$, we define
\begin{equation}
N_o=\displaystyle{\sum_{u\in\mathcal{U}}}\mathbf{1}\lbrace\mathbb{E}[T_u\mid \mathsf{P},\mathsf{F},\mathsf{G}]=0\rbrace
\end{equation}
as the number of users that in outage, where $\mathbf{1}\lbrace\mathbb{E}[T_u\mid \mathsf{P},\mathsf{F},\mathsf{G}]=0\rbrace$ is the indicator function such that the value is $1$ if $\mathbb{E}[T_u\mid \mathsf{P},\mathsf{F},\mathsf{G}]=0$; otherwise the value is $0$. Intuitively, $\mathbf{1}\lbrace\mathbb{E}[T_u\mid \mathsf{P},\mathsf{F},\mathsf{G}]=0\rbrace$ is equal to zero when the file delivery scheme cannot deliver the desired file to user $u$. We note that the expectation of $\mathbb{E}[T_u\mid \mathsf{P},\mathsf{F},\mathsf{G}]$ is taken over the file delivery scheme. The outage probability is then defined as
\begin{equation}
p_o=\frac{1}{N}\mathbb{E}_{\mathsf{P},\mathsf{F},\mathsf{G}}[N_o]=\frac{1}{N}\sum_{u\in\mathcal{U}}\mathbb{P}\left(\mathbb{E}[T_u\mid \mathsf{P},\mathsf{F},\mathsf{G}]=0\right).
\end{equation}

\subsection{Channel Model}

We consider the generalized physical model in this paper. Suppose there is a TX-RX pair $u$, where user $u$ serves as the TX and user $u^{(\text{r})}$ serves as RX. We denote $x_u$ and $x_{u^{(\text{r})}}$ as the locations of user $u$ and $u^{(\text{r})}$, respectively, and denote $\Gamma_{\text{Co}}^u$ as the set of users transmitting in the same time-frequency resource. Assume that $P_{\text{max}}$ is the maximum power that a user can use for transmission. Then, the generalized physical model defines the link-rate of the TX-RX pair $u$ as \cite{agarwal2004capacity,xue2006scaling}:
\begin{equation}\label{eq:gen_model}
R(u,u^{(\text{r})})=B_u\log_2\left(1+\frac{P_ul_{uu^{(\text{r})}}}{B_uN_0+\sum_{k\neq u,k\in\Gamma^u_{\text{Co}}}P_kl_{ku^{(\text{r})}}}\right),
\end{equation}
where $B_{u}$ is the bandwidth used for communication between users $u$ and $u^{(\text{r})}$; $P_u\leq P_{\text{max}}$ is the power of user $u$; and $l_{uu^{(\text{r})}}=\frac{\chi}{\left(d_{uu^{(\text{r})}}\right)^{\alpha}}$ is the path (power) gain between users $u$ and $u^{(\text{r})}$,\footnote{It should be noted that the adopted path gain model is an approximation of the realistic model, as it could violate the rationale that the transmit power is larger than the receive power in certain regime and that the pathloss coefficient should follow the free-space pathloss model in the near-field regime. Nevertheless, the adopted model can appropriately capture the relative power loss among different TX-RX pairs and provide high tractability. Therefore, this model is effective and useful as we can correctly interpret the derived results in consideration of the potentially unreasonable portions of the results which will be discussed in more detail later in this paper.} where
\begin{equation}
d_{uu^{(\text{r})}}=\vert x_u-x_{u^{(\text{r})}}\vert
\end{equation}
is the distance between users $u$ and $u^{(\text{r})}$, $\chi>0$ is some calibration factor, and $\alpha> 2$ is the pathloss coefficient. Note that different from our conference version that provides dedicated analysis for both the bounded physical model and the generalized physical model, in this paper, we focus on the analysis of the generalized physical model. This is because the scaling laws derived in consideration of the bounded physical model can be treated as special cases for those derived for the generalized physical model (with the assumption that $P_u$ is finite). Specifically, as indicated in \cite{agarwal2004capacity,xue2006scaling}, for any configuration of TX-RX pairs, the differences between the link-rates of TX-RX pairs considering these two physical models are simply bounded by some finite constant when $P_u,\forall u$ are finite. Therefore, it is sufficient that we focus on the generalized physical model. Note that to obtain the rigourous derivations for the scaling laws with the bounded physical model, we can repeat the derivations in this paper and apply them to the bounded physical model after some modifications similar to the approach provided in \cite{lee2021throughput}.

\subsection{Targeting Scenarios}

We in this paper consider two scenarios with different assumptions. Specifically, for {\em each realization of the network}, we assume (in the order-wise sense) either (i) all TX-RX pairs transmit the same number of bits in $T'$ sec (e.g., users obtain the same amount of segments of files in $T'$ sec) or (ii) different TX-RX pairs can transmit different numbers of bits in $T'$ sec. Since the first assumption forces different TX-RX pairs in a network realization to have equal user throughput, we refer it to as ``equal-throughput assumption''. On the other hand, when the equal-throughput assumption is relaxed, i.e., when considering the second assumption, different TX-RX pairs of a network realization can have different throughput, indicating receptions of more bits are allowed for some users in the network realization. Note that although the second assumption would lead to $T_u\neq T_v$ for some $u\neq v$, we still have $\overline{T}_{u}=\overline{T}_{v},\forall u,v$ due to symmetry of the network. Hence, the unequal-throughput assumption considered in the second scenario is in the per realization sense, instead of the average sense. This implies that the unfairness happening for the second scenario is only in the realization level, and users are still statistically fair so that the optimal throughput with definition in (\ref{eq:throu_def}) is non-trivial.

In the remainder of this paper, we will analyze the network with and without the equal-throughput assumption, where the first scenario, i.e., the scenario with the equal-throughput assumption, is denoted as scenario 1 and the second scenario is denoted as scenario 2.

\section{Throughput-Outage Analysis for MZipf Distribution with $\gamma<1$}

\label{Sec:Out_Bound_Gen}

In this section, we analyze the throughput-outage performance for the case $\gamma<1$. We will first provide the outer bound analysis for both scenarios introduced in Sec. \ref{Sec:Net_Setup}.B. Then, the achievable schemes and the analyses corresponding to each scenario are provided. We start the analysis by providing Lemmas 1 and 2 which characterize the outage probability and by providing Theorem 1 which describes the transport capacity upper bound.

{\em Lemma 1 (Lemma 4 in\cite{lee2020optimal}):} When $n=\omega(M)$ users are uniformly distributed within a network with unit size, the probability to have $N_{\text{D}}$ users within an area of size $A=o\left(\frac{N_{\text{D}}}{n}\right)$ is upper bounded by $o(1)$.

{\em Lemma 2 (Lemma 5 in\cite{lee2020optimal}):} Suppose $\gamma<1$. Then, when a user in the network searches through $n_{\text{s}}=o\left(\frac{M}{S}\right)$ different users, we obtain $p_{\text{miss}}(n_{\text{s}})\geq 1-o(1)$, where $p_{\text{miss}}(n_{\text{s}})$ is the probability that the user cannot find the desired file from the $n_{\text{s}}$ users. Furthermore, when a user in the network searches through $n_{\text{s}}=\rho' M$ different users for some $\rho'$, we have the following results: (i) $p_{\text{miss}}(n_{\text{s}})\geq \Theta\left(e^{-\rho'}\right)$ if $\rho'=\Theta(1)$ is large enough; and (ii) $p_{\text{miss}}(n_{\text{s}})\geq (1-\gamma)e^{-(S\rho'-\gamma)}$ if $\rho'=\omega(1)$.

{\em Theorem 1:} We denote the set of TX-RX pairs as $\Gamma$ and define $r_u$ as the communication distance for the TX-RX pair $u$. Let $M\to\infty$, $N\to\infty$, and $q\to\infty$. Suppose $\gamma<1$ and $q=\mathcal{O}(M)$. We let $R_0=\epsilon_0\sqrt{\frac{\rho'M}{SN}}$, where $\epsilon_0$ is some small constant. We denote the transport capacity of the network consisting of $\Gamma$, defined in terms of meter-bits/s, as $C_{\Gamma}=\sum_{u\in\Gamma}r_uC_u$, where $C_u$ is the average rate (bits/s) of user $u$. We denote the set of TX-RX pairs which have the largest powers among the TX-RX pairs in their corresponding time-frequency resources as $\mathcal{W}$. 
Under the generalized physical model, 
$C_{\Gamma}$ is upper bounded as:
\begin{equation}
\begin{aligned}\label{eq:g_model_TC_Thm}
C_{\Gamma}&\leq B\overline{C}_{\mathcal{W}}+B\overline{C}_{\Gamma_{R_0}}+B\frac{\log_2(e)}{\epsilon_0}\sqrt{\frac{SN}{\rho'M}}\left(\alpha\left(3\sqrt{2}+1\right)+2(2(\sqrt{2}+1))^{\alpha}\right),
\end{aligned}
\end{equation}
where $B$ is the total bandwidth of the network; $\overline{C}_{\mathcal{W}}$ is the average transport capacity efficiency, defined in terms of meter-bits per second per Hz, of the TX-RX pairs in $\mathcal{W}$; $\overline{C}_{\Gamma_{R_0}}$ is the average transport capacity efficiency of TX-RX pairs that are not in $\mathcal{W}$ and have communication distances smaller than $R_0$. The definitions of $\overline{C}_{\mathcal{W}}$ and $\overline{C}_{\Gamma_{R_0}}$ are formally given below (\ref{eq:g_model_TC}) at the end of Appendix \ref{App:ProofThm1}.

\begin{proof}
See Appendix \ref{App:ProofThm1}.
\end{proof}

{\em Remark 1:} From Lemmas 1 and 2, we conclude that to have a non-vanishing probability for a user to obtain the desired file (i.e., $p_{\text{miss}}(n)$ does not go to $1$), with high probability, the distance between the TX and RX is at least $\Theta\left(\sqrt{\frac{\rho'M}{SN}}\right)$, where $\rho'=\Omega(1)$.

\subsection{Outer Bound Analysis for Scenario 1} 

Here, we consider scenario 1 and provide the outer bound. Since the equal-throughput assumption is considered, different TX-RX pairs transmit the same number of bits in a time period of $T'$ sec. With Remark 1 and Theorem 1, we can obtain the following theorem:

{\em Theorem 2:} Let $M\to\infty$, $N\to\infty$, and $q\to\infty$. Suppose $\gamma<1$. Assume that the powers of users in the network are upper bounded by $P_{\text{max}}$. When $\rho'=\Omega(1)$ is large enough, the throughput-outage performance of the network is dominated by:
\begin{equation}
\begin{aligned}
T(P_o)&= \Theta\left(\frac{B}{N}\log_2\left(1+\frac{P_{\text{max}}}{N_0B_{\text{s}}}\frac{\chi}{\left(\frac{\rho'M}{SN}\right)^\frac{\alpha}{2}}\right)+B\log_2(e)\left(\alpha\left(3\sqrt{2}+1\right)+2(2(\sqrt{2}+1))^{\alpha}\right)\frac{S}{\rho'M}\right),\\
P_o&= \Theta\left(e^{-\rho'}\right),
\end{aligned}
\end{equation}
where $P_o$ can be arbitrarily small or converging to zero.
\begin{proof}
See Appendix \ref{App:ProofThm2}.
\end{proof}

{\em Remark 2:} Note that when $N\to\infty$, $T(P_o)$ can be unbounded because the term $\frac{\chi}{\left(\frac{\rho'M}{SN}\right)^\frac{\alpha}{2}}$ in $T(P_o)$ in Theorem 2 can go to infinity. This unreasonable result is brought by having the cases that the signal-to-interference-plus noise ratio (SINR) becomes unbounded due to the unbounded path gain. Thus, to correctly interpret the result in Theorem 2, we should consider that the term $\frac{\chi}{\left(\frac{\rho'M}{SN}\right)^\frac{\alpha}{2}}$ in $T(P_o)$ in Theorem 2 is upper bounded by $1$, which corresponds to the physical reality of the pathloss laws. Note that such consideration applies to all results in the remainder of this paper.

{\em Remark 3:} Theorem 2 shows that when $\gamma<1$, $M$ is the dominant factor while $q$ does not impact the asymptotic scaling law. In addition, it shows that when the maximum transmit power is some constant, the throughput-outage performance outer bound considering the generalized physical model has the same scaling law as the throughput-outage performance considering the protocol model \cite{Ji:Th_Out_toff,lee2019throughput}. Note that in contrast to the physical model here which enables the link-level power allocation and scheduling, the protocol model and the approaches in \cite{Ji:Th_Out_toff,lee2019throughput} only consider the simple clustering network and the system-level changing of the cluster size. As a result, this indicates that the link-level power allocation and scheduling cannot improve the throughput-outage scaling law, i.e., the asymptotic growth rate of the throughput-outage performance, when requests of users are served with equal-throughput assumption. That being said, in practice, the constant factor of the throughput-outage performance might still be improved by a good power control and link scheduling approach.

{\em Remark 4:} Slightly different from Remark 3, when we allow the maximum instantaneous transmit power to be infinity while the average power is still some constant, Theorem 2 suggests that the asymptotic performance might be improved if $\frac{S}{\rho'M}=o\left(\frac{\log_2(N)}{N}\right)$. Such improvement could be possible if we let a user to exclusively transmit with the power being $\Theta(N)$ once every $\Theta(N)$ time-slots. However, in this case, it indicates that simple time-division multiple access (TDMA) can dominate the performance, and thus the relevant discussion becomes trivial.

\subsection{Outer Bound Analysis for the Proposed Throughput-Enhancing Approach in Scenario 2} 

From the result in Sec. \ref{Sec:Out_Bound_Gen}.A, we see that having link-level power allocation and scheduling cannot improve the throughput-outage scaling law when the equal-throughput assumption is considered. To break such limitation, we drop the equal-throughput assumption here and propose a throughput-enhancing approach that can improve the scaling law by having link-level power allocation and scheduling. We conduct the outer bound analysis for the proposed throughput-enhancing approach in this subsection. The achievable performance for the throughput-enhancing approach will then be discussed later in Sec. \ref{Sec:Out_Bound_Gen}.D.

Since the equal-throughput assumption is dropped in this case, we can take advantage of letting the TX-RX pairs with small communication distances transmit at a much higher rate to enhance the network throughput. {Based on this concept, we propose a double time-slot throughput-enhancing approach as follows.} We first split the overall transmission period $T'$ into two time-slots; each has the duration $\frac{T'}{2}$. The first time-slot is used for the general file delivery which adopts the same approach as in scenario 1. We then use the second time-slot to enhance the overall throughput. To do this, we let TX-RX pairs with communication distances smaller than $\sqrt{\epsilon'}R_0$, where $\epsilon'=\mathcal{O}(1)$, transmit with high throughput in the second time-slot. Note that different users in the second time-slot are assumed to transmit with equal throughput, but such throughput should be much larger than the throughput of users transmitting in the first time-slot. In addition, the split of the overall transmission period into two equal time-slots does not lose the optimality of the scaling law as compared to other time-slot splits using different fractions. This is because the scaling law is to characterize the order gain, and we only have a constant factor gain even if we can magically allow the transmission durations for both time-slots to be extended to $T'$, i.e., both time-slots occupy a duration of $T'$ (which is not possible in reality as the overall transmission period is only $T'$). With the above described approach, we can obtain the following theorem that characterizes the outer bound for the proposed double time-slot scheme:

{\em Theorem 3:} Let $M\to\infty$, $N\to\infty$, and $q\to\infty$. Assume $\gamma<1$ and $q=\mathcal{O}\left(\epsilon'\rho'\frac{M}{S}\right)$. Suppose the double time-slot framework discussed in Sec. \ref{Sec:Out_Bound_Gen}.B is used and the $\epsilon'=\mathcal{O}(1)$ is selected. Assume that the powers of users in the network are upper bounded by $P_{\text{max}}$. Then, when $\rho'=\Omega(1)$ is large enough and $\lambda_2$ is feasible, the throughput-outage performance of the network is dominated by:
\begin{equation}
\begin{aligned}\label{eq:g_model_TC_Anal_Final_C2}
T(P_o)=\Theta\left(\frac{B}{2}\frac{\log_2\left(1+\frac{P_{\text{max}}}{N_0B_{\text{s}}}\frac{\chi}{\left(\frac{\epsilon'\rho'M}{SN}\right)^\frac{\alpha}{2}}\right)}{N}+\frac{B}{2}\frac{S}{\epsilon'\rho'M}\right),P_o= \Theta\left(e^{-\rho'}\right).
\end{aligned}
\end{equation}
Furthermore, when considering a {\em network instance}, the throughput per user for users with communication distances $d_u\geq \sqrt{\epsilon'}R_0$ is dominated by:
\begin{equation}
\lambda_1= \Theta\left(\frac{B}{N}\log_2\left(1+\frac{P_{\text{max}}}{N_0B_{\text{s}}}\frac{\chi}{\left(\frac{\rho'M}{SN}\right)^\frac{\alpha}{2}}\right)+\frac{BS}{\rho'M}\right);
\end{equation}
the throughput per user for users with communication distance $d_u\leq\sqrt{\epsilon'}R_0$ is dominated by:
\begin{equation}
\lambda_2=\Theta\left(\frac{B}{2}\frac{\log_2\left(1+\frac{P_{\text{max}}}{N_0B_{\text{s}}}\frac{\chi}{\left(\frac{\epsilon'\rho'M}{SN}\right)^\frac{\alpha}{2}}\right)}{\delta'N}\right)+\Theta\left(\frac{BS}{\delta'\epsilon'\rho'M}\right),
\end{equation}
where $\delta'N$ is the number of users with communication distance $d_u\leq\sqrt{\epsilon'}R_0$.

\begin{proof}
See Appendix \ref{App:ProofThm3}.
\end{proof}

{\em Remark 5:} By Theorem 3, we see that the concept of leveraging high-throughput transmissions of the TX-RX pairs with small distances can effectively enhance the overall throughput outer bound as our proposed double time-slot approach is realized with $\epsilon'=o(1)$. We stress that this is an outer bound for the proposed double time-slot scheme, while we make no claims about outer bounds for all possible transmission schemes. Therefore, it is likely that introducing the multiple (more than two) time-slot approach might further enhance the throughput. However, as the performance enhancement ability is dependent on $\epsilon'$ and $\delta'$, and the characteristics of such performance enhancement ability are unclear, we thus in this paper focus on studying the performance of the double time-slot framework, and the investigations of the multiple time-slot framework are considered as possible future works.

We see from Theorem 3 that the network throughput $T$ can be increased via decreasing $\epsilon'$. However, due to the physical limitation, namely the TX-RX link feasibility, there is a lower bound on $\epsilon'$, leading to an upper bound of the throughput. To find this upper bound, we in the following analyze $\epsilon'$. Note that the first term of (\ref{eq:g_model_TC_Anal_Final_C2}) is asymptotically irrelevant to $\epsilon'$ because we interpret $\frac{\chi}{\left(\frac{\epsilon'\rho'M}{SN}\right)^\frac{\alpha}{2}}$ is upper bounded by some constant according to Remark 2. Thus, the benefit of the first term of (\ref{eq:g_model_TC_Anal_Final_C2}) comes only from letting the transmission power go to infinity. It follows that we can without loss of generality focus on the $\frac{B}{2}\frac{S}{\epsilon'\rho'M}$ term when characterizing the lower bound of $\epsilon'$. We thus in the following assume that $P_{\text{max}}$ is some constant for simplicity, and then derive the following Corollary:

{\em Corollary 1:} Following Theorem 3, when maximizing the number of users that can obtain the desired files within the distance $\sqrt{\epsilon'}R_0$, namely when maximizing $\delta'$, the throughput for users with communication distances being $d_u\leq\sqrt{\epsilon'}R_0$ is dominated by:
\begin{equation}
\lambda_2=\Theta\left(\frac{B}{2}\frac{\log_2\left(1+\frac{P_{\text{max}}}{N_0B_{\text{s}}}\frac{\chi}{\left(\frac{\epsilon'\rho'M}{SN}\right)^\frac{\alpha}{2}}\right)}{(\epsilon'\rho')^{1-\gamma}N}\right)+\Theta\left(\frac{BS}{(\epsilon'\rho')^{2-\gamma}M}\right).
\end{equation}
\begin{proof}
See Appendix \ref{App:ProofCoro1}.
\end{proof}

With Corollary 1, we can then derive the outer bound for the proposed throughput-enhancing approach. This is elaborated as follows. By using the same analysis as that for scenario 1, we first know that the throughput per user in the second time-slot leads to the following upper bound (see Appendix \ref{App:ProofThm2} for details):
\begin{equation}\label{eq:out_lambda_2xx}
\lambda_2\delta' N\Theta\left(\sqrt{\frac{\epsilon'\rho'M}{SN}}\right)=\mathcal{O}\left(B\log_2\left(1+\frac{P_{\text{max}}}{N_0B_{\text{s}}}\frac{\chi}{\left(\frac{\epsilon'\rho'M}{SN}\right)^\frac{\alpha}{2}}\right)\sqrt{\frac{\epsilon'\rho'M}{SN}}+B\sqrt{\frac{SN}{\epsilon'\rho'M}}\right).
\end{equation} 
With (\ref{eq:out_lambda_2xx}) and that $P_{\text{max}}$ is some constant, we obtain
\begin{equation}\label{eq:delta_bound_SecIIIA_2}
\lambda_2\delta'=\mathcal{O}\left(\frac{S}{\epsilon'\rho'M}\right).
\end{equation}
Then, observe that if we want $\lambda_2$ to be less likely to hit its upper bound for a given $\epsilon'$, we shall maximize $\delta'$. Recall that the maximum is $\delta'=\Theta\left((\epsilon'\rho')^{1-\gamma}\right)$ as indicated in Corollary 1. This along with that the TX-RX feasibility condition to satisfy is $\lambda_2\leq \eta$, where $\eta$ is some constant indicating that the link-rate cannot be infinitely large, leads to that if we want to minimize $\epsilon'$ while maintaining the tightness of the upper bound, we should have
\begin{equation}
\eta(\epsilon'\rho')^{1-\gamma}=\Theta\left(\frac{S}{\epsilon'\rho'M}\right)
\end{equation}
Since $\eta$ is some constant, this then lead to
\begin{equation}\label{eq:epislon_bound_SecIIIA}
\epsilon'\rho'=\Theta\left(\left(\frac{S}{M}\right)^{\frac{1}{2-\gamma}}\right).
\end{equation}
Finally, by using Theorem 3, Corollary 1, and (\ref{eq:epislon_bound_SecIIIA}), we obtain Theorem 4 as following:

{\em Theorem 4:} Let $M\to\infty$, $N\to\infty$, and $q\to\infty$. Assume $\gamma<1$ and $q=\mathcal{O}\left(\left(\frac{M}{S}\right)^{\frac{1-\gamma}{2-\gamma}}\right)$. Suppose the double time-slot framework discussed in Sec. \ref{Sec:Out_Bound_Gen}.B is used and the $\epsilon'=\mathcal{O}(1)$ is selected. Assume that the powers of users in the network are upper bounded by $P_{\text{max}}$. When $\rho'=\Omega(1)$ is large enough, the throughput-outage performance of the network is dominated by:
\begin{equation}
\begin{aligned}
T=\Theta\left(\frac{B}{2}\frac{\log_2\left(1+\frac{P_{\text{max}}}{N_0B_{\text{s}}}\frac{\chi}{\left(\left(\frac{M}{S}\right)^{\frac{1-\gamma}{2-\gamma}}\frac{1}{N}\right)^\frac{\alpha}{2}}\right)}{N}+\frac{B}{2}\left(\frac{S}{M}\right)^{\frac{1-\gamma}{2-\gamma}}\right),P_o= \Theta\left(e^{-\rho'}\right).
\end{aligned}
\end{equation}
Furthermore, when considering a {\em network instance}, the throughput per user for users with communication distances $d_u\geq \sqrt{\epsilon'}R_0$ is dominated by:
\begin{equation}
\lambda_1= \Theta\left(\frac{B}{N}\log_2\left(1+\frac{P_{\text{max}}}{N_0B_{\text{s}}}\frac{\chi}{\left(\frac{\rho'M}{SN}\right)^\frac{\alpha}{2}}\right)+B\log_2(e)\left(\alpha\left(3\sqrt{2}+1\right)+2(2(\sqrt{2}+1))^{\alpha}\right)\frac{S}{\rho'M}\right);
\end{equation}
the throughput per user for users with communication distances $d_u\leq\sqrt{\epsilon'}R_0$ is dominated by:
\begin{equation}
\lambda_2=\Theta\left(\frac{B}{2}\frac{\log_2\left(1+\frac{P_{\text{max}}}{N_0B_{\text{s}}}\frac{\chi}{\left(\left(\frac{M}{S}\right)^{\frac{1-\gamma}{2-\gamma}}\frac{1}{N}\right)^\frac{\alpha}{2}}\right)}{\left(\left(\frac{S}{M}\right)^{\frac{1-\gamma}{2-\gamma}}\right)N}\right)+\Theta(B).
\end{equation}
\begin{proof}
This is directly obtained by using Theorem 3, Corollary 1, and (\ref{eq:epislon_bound_SecIIIA}).
\end{proof}

{\em Remark 6:} By comparing between Theorem 2 and Theorem 4, we observe that the double time-slot approach can significantly improve the throughput performance without sacrificing the outage probability. The benefit is from that we judiciously let TX-RX pairs with very small communication distances transmit at a much higher throughput in the second time-slot. Note that here we implicitly assume that users have sufficient demands so that the only limitation for a user to increase its throughput is the link-rate. In addition, we see that $T_2=\Theta(B)$ is some constant if the maximum transmit power is some constant. This indicates that the TX-RX pairs with enhanced throughput satisfy the link-rate feasibility condition.

{\em Remark 7:} From Theorem 4, we observe that using the double time-slot approach indeed gives rise to some degree of unfairness, as for each network realization, the users allowed to transmit in the second time-slot can enjoy a much higher instantaneous throughput, though different users would have the same average throughput. Furthermore, as indicated by Theorem 4 that the network throughput $T$ is independent of $P_o$, the double time-slot framework can ultimately decouple the tradeoff between throughput and outage, and thus the throughput-outage tradeoff no longer exists in terms of the average user throughput. However, this is because the TX-RX pairs with small communication distances can maintain the overall network throughput when increasing $\rho'$ at the cost of introducing further unfairness, and thus the overall throughput-outage tradeoff has been converted to fairness-outage tradeoff in this context. Finally, we stress that the throughput-enhancing result here is not because we let TX-RX pairs with high throughput to transmit the same amount of information bits as those TX-RX pairs with low throughput, and then let them finish their transmissions fast so that the remaining amount of resource for other TX-RX pairs to transmit can be increased. On the other hand, the overall throughput is enhanced because the TX-RX pairs with high throughput indeed successfully transmit and receive much more information bits than those TX-RX pairs with low throughput in a given time period.

\subsection{Achievable Scheme and Analysis for Scenario 1}

\label{Sec:Ach_gl1_C1}

Here, we provide the achievable scheme and its corresponding analysis for scenario 1. We consider the following achievable scheme. Suppose the communications are in $T'$ sec. We first split this $T'$-second period into two time-slots; each has $\frac{T'}{2}$ sec. Then, in the first time-slot, all $N$ users in the network are served in a round-robin manner using time division multiple access (TDMA) approach, in which each of them can transmit with the maximum power $P_{\text{max}}$ in a period of $\frac{T'}{2N}$ sec. In the second time-slot, we adopt the clustering network with the frequency reuse scheme and cluster-wise round-robin scheduling, i.e., users in the same cluster are served in the round-robin manner and different clusters can be activated simultaneously. The clustering approach used in the second time-slot is as follows. We split the network into equally-sized square clusters whose side length is $d=\Theta\left(\sqrt{\frac{\rho M}{SN}}\right)$. We assume users in a cluster can obtain the desired file only from users in the same cluster. In each cluster, a user is served at a time and users in the same cluster are served in a round-robin manner. Interference between different clusters is avoided via the frequency reuse approach with reuse factor $(2(K+1))^2$, where $K\in\mathcal{N}>0$ is a finite positive integer. Then, by symmetry of the clusters and by the facts that $N\to\infty$ and $d=o(1)$, when computing the outage probability, we can assume that clusters are independent to one another and that the number of users $N_{\text{cluster}}$ in a cluster follows the Poisson distribution, given as $N_{\text{cluster}}\backsim F_{\text{Poi}}\left(\frac{\rho M}{S}\right)$, where $\frac{\rho M}{S}$ is the mean value of the Poisson distribution. Note that the latter assumption is because the binomial distribution converges to a Poisson distribution when the number of trials goes to infinity while the probability of success goes to zero. 

We use the randomized caching policy described in Sec. \ref{Sec:Net_Setup}. It follows from Lemma 1 in \cite{lee2020optimal} that the outage probability of the described network with clustering is 
\begin{equation}\label{eq:Ach_min_Prob}
p_o=\sum_{f=1}^M P_r(f) e^{-\frac{\rho M}{S}P_c(f)},
\end{equation}
where $P_c(f),\forall f$ are determined by the caching policy minimizing (\ref{eq:Ach_min_Prob}), as provided in Theorem 1 of \cite{lee2020optimal}. With the aforementioned transmission and caching policies, users in the network are served by the combination of two types of delivery approaches, i.e., the simple TDMA in the first time-slot and the clustering in the second time-slot. Note that the split here is not to enable the throughput-enhancing approach introduced in Sec. \ref{Sec:Out_Bound_Gen}.B as scenario 1 is considered here. On the contrary, it is simply used for achieving the outer bound in Sec. \ref{Sec:Out_Bound_Gen}.A. To derive the achievable throughput-outage performance, we start with the following proposition:

{\em Proposition 1:} Suppose the clustering network is considered with the frequency reuse approach adopting the reuse factor $(2(K+1))^2$, where $K\in\mathcal{N}>0$ is some finite positive integer. Assume that the powers of users in the network are upper bounded by some constant $\nu_{\text{upp}}=\Theta(1)$ and lower bounded by some constant $\nu_{\text{low}}=\Theta(1)$, i.e., $\nu_{\text{low}}\leq P_u\leq \nu_{\text{upp}},\forall u$. Then, when each cluster at most activates a single user in the cluster for transmission at a time, there must exist some constant $\vartheta$ such that for any activated TX-RX pair $u$, we obtain
\begin{equation}
R(u,u^{(\text{r})})=B_u\log_2\left(1+\frac{P_ul_{uu^{(\text{r})}}}{B_uN_0+\sum_{k\neq u,k\in\Gamma^u_{\text{Co}}}P_kl_{ku^{(\text{r})}}}\right)\geq B_u\log_2(1+\vartheta),
\end{equation}
where $B_u=\frac{B}{(2(K+1))^2}$ and $\vartheta$ is monotonically increasing with respect to the reuse factor $K$.
\begin{proof}
See Appendix \ref{App:ProofProp1}.
\end{proof}

Proposition 1 indicates that by using the proposed clustering with frequency reuse scheme, users in different clusters are guaranteed to have some constant link-rate. Then, by combining the transmissions in the first and second time-slots and leveraging Proposition 1, we obtain the following theorem:

{\em Theorem 5:} Let $M\to\infty$, $N\to\infty$, and $q\to\infty$. Suppose $\gamma<1$ and $q=\mathcal{O}(M)$. Assume that the powers of users in the network are upper bounded by $P_{\text{max}}$. Consider the caching policy in Theorem 1 of \cite{lee2020optimal} and that the side length of a cluster is $\sqrt{\frac{\rho M}{SN}}$. Assume equal-throughput transmissions of users. When $\rho=\Omega(1)$ is large enough, the following throughput-outage performance of the network is achievable:
\begin{equation}
\begin{aligned}
&T(P_o)= \Theta\left(\frac{B}{N}\log_2\left(1+\frac{P_{\text{max}}}{N_0B}\frac{\chi}{\left(\frac{\rho M}{SN}\right)^\frac{\alpha}{2}}\right)\right)+\Theta\left(\frac{BS}{\rho M}\right),P_o= \Theta\left(e^{-\rho}\right),
\end{aligned}
\end{equation}
where $P_o$ can be negligibly small or converging to zero.
\begin{proof}
See Appendix \ref{App:ProofThm5}.
\end{proof}

{\em Remark 8:} By comparing between Theorems 2 and 5, we see that the proposed outer bound is achievable. This indicates that when the equal-throughput assumption is considered, the simple clustering scheme is asymptotically optimal even though we are allowed to use the link-level power allocation and scheduling.

\subsection{Achievable Scheme and Analysis for Scenario 2 with the Proposed Throughput-Enhancing Approach}

Here, an achievable scheme for scenario 2 is presented and analyzed. The achievable scheme for scenario 2 is a combination of the achievable scheme for scenario 1 and the double time-slot throughput-enhancing approach introduced in Sec. \ref{Sec:Out_Bound_Gen}.B. Thus, for the achievable scheme here, by following the throughput-enhancing approach, we first split the transmission duration $T'$ sec into two time-slots; each has $\frac{T'}{2}$ sec. We assume without loss of generality that $S$ is an even number. Then, in the first time-slot, the achievable scheme proposed for scenario 1 in Sec. \ref{Sec:Out_Bound_Gen}.C is directly used, where cluster size in this case is set to $d_1=\sqrt{\frac{\rho M}{SN}}$. In the second time-slot, we also consider the achievable scheme proposed for scenario 1 in Sec. \ref{Sec:Out_Bound_Gen}.C. However, in this case, the side length of the cluster is changed to $d_2=\sqrt{\frac{\epsilon\rho M}{SN}}$, where $\epsilon=\mathcal{O}(1)$. By the above descriptions, we observe that the achievable scheme in Sec. \ref{Sec:Out_Bound_Gen}.C is used twice, and each has a dedicated cluster size for realizing the clustering approach used in each time-slot.

For the caching scheme, we also split the whole cache space into two subspaces, and each has size $\frac{S}{2}$. For the first caching subspace, we consider the caching policy proposed in Theorem 1 of \cite{lee2020optimal} with $g_{c,1}(M)=\frac{2\rho M}{S}$. For the second caching subspace, we consider again the same caching policy and let $g_{c,2}(M)=\frac{2\epsilon\rho M}{S}$. By the above described transmission and caching policies, we can then obtain the following theorem which characterizes the achievable performance:

{\em Theorem 6:} Let $M\to\infty$, $N\to\infty$, and $q\to\infty$. Suppose $\gamma<1$ and $q=\mathcal{O}\left(\left(\frac{M}{S}\right)^{\frac{1-\gamma}{2-\gamma}}\right)$. Suppose the proposed achievable scheme in Sec. \ref{Sec:Out_Bound_Gen}.D is used and $\epsilon$ is selected such that $\epsilon\rho=\Theta\left(\left(\frac{S}{M}\right)^{\frac{1}{2-\gamma}}\right)$. Assume that the powers of users in the network are upper bounded by $P_{\text{max}}$. Then, when $\rho=\Omega(1)$ is large enough, the following throughput-outage performance of the network is achievable:
\begin{equation}
\begin{aligned}
T(P_o)=\Theta\left(\frac{B}{2}\frac{\log_2\left(1+\frac{P_{\text{max}}}{N_0B_{\text{s}}}\frac{\chi}{\left(\left(\frac{M}{S}\right)^{\frac{1-\gamma}{2-\gamma}}\frac{1}{N}\right)^\frac{\alpha}{2}}\right)}{N}+\frac{B}{2}\left(\frac{S}{M}\right)^{\frac{1-\gamma}{2-\gamma}}\right),P_o= \Theta\left(e^{-\rho}\right).
\end{aligned}
\end{equation}
Furthermore, when considering a {\em network instance}, the achievable throughput per user for users with communication distances $d_u\geq \sqrt{\frac{\epsilon\rho M}{SN}}$ is:
\begin{equation}
T_1= \Theta\left(\frac{B}{N}\log_2\left(1+\frac{P_{\text{max}}}{N_0B_{\text{s}}}\frac{\chi}{\left(\frac{\rho M}{SN}\right)^\frac{\alpha}{2}}\right)+\frac{BS}{\rho M}\right);
\end{equation}
the achievable throughput per user for users with communication distances $d_u<\sqrt{\frac{\epsilon\rho M}{SN}}$ is:
\begin{equation}
T_2=\Theta\left(\frac{B}{2}\frac{\log_2\left(1+\frac{P_{\text{max}}}{N_0B_{\text{s}}}\frac{\chi}{\left(\left(\frac{M}{S}\right)^{\frac{1-\gamma}{2-\gamma}}\frac{1}{N}\right)^\frac{\alpha}{2}}\right)}{\left(\left(\frac{S}{M}\right)^{\frac{1-\gamma}{2-\gamma}}\right)N}\right)+\Theta(B).
\end{equation}
\begin{proof}
See Appendix \ref{App:ProofThm6}.
\end{proof}

{\em Remark 9:} By comparing between Theorems 4 and 6, we see that the proposed outer bound is achievable. However, different from scenario 1 where the optimality can be achieved without resorting to link-level power control and scheduling, the achievable scheme of scenario 2 exploits the link-level power control and scheduling to enhance the throughput of users in the second time-slot so that the overall throughput is increased. Note that the use of different cluster sizes for different time-slots implies that power control and scheduling are used in link-level such that TX-RX pairs scheduled in different time-slots can follow the required cluster size and scheduling. This indicates that the link-level power control and scheduling can significantly improve the network throughput at the cost of some degree of fairness.

\section{Throughput-Outage Analysis for MZipf Distribution with $\gamma>1$}

\label{Sec:Out_Bound_Gen_gg1}

In this section, we analyze the throughput-outage performance considering $\gamma>1$ and $q=o(M)$. Similar to Sec. III, we will in this section first derive the outer bounds for scenario 1 and scenario 2 with the throughput-enhancing approach, and then provide the achievable schemes along with the corresponding throughput-outage performance analysis. We start the analysis by providing Lemma 3 and Theorem 7 which characterize the outage probability lower bound and the transport capacity upper bound, respectively.

{\em Lemma 3 (Lemma 8 in\cite{lee2020optimal}):} Suppose $\gamma>1$. Considering $q=o(M)$, we have the following results: (i) when a user searches through $n_{\text{s}}=o\left(\frac{q}{S}\right)$ different users in the network, we obtain $p_{\text{miss}}(n)\geq 1-o(1)$; and (ii) when a user searches through $n_{\text{s}}=\frac{\alpha_1' q}{S}<\frac{M}{S}$ different users, where $\alpha_1'=\Omega\left(1\right)$ but $\alpha_1'=\mathcal{O}\left(q^{\frac{1}{\gamma-1}}\right)$, we obtain $p_{\text{miss}}(n)\geq\Theta\left(\frac{1}{(\alpha_1')^{\gamma-1}}\right)$.

{\em Theorem 7:} We denote the set of TX-RX pairs as $\Gamma$ and define $r_u$ as the communication distance for the TX-RX pair $u$. Let $M\to\infty$ and $N\to\infty$. Suppose $\gamma>1$ and $\alpha_1'q=o(M)$. We let $R_0'=\epsilon_0\sqrt{\frac{\alpha_1'q}{SN}}$, where $\epsilon_0$ is some small constant. We denote the transport capacity of the network consisting of $\Gamma$, defined in terms of meter-bits/s, as $C_{\Gamma}=\sum_{u\in\Gamma}r_uC_u$, where $C_u$ is the average throughput (bits/s) of user $u$. We denote the set of TX-RX pairs which have the largest powers among the TX-RX pairs in their corresponding time-frequency resources as $\mathcal{W}$. Under the generalized physical model, $C_{\Gamma}$ is upper bounded as:
\begin{equation}
\begin{aligned}\label{eq:g_model_TC_Thm_gg1}
C_{\Gamma}&\leq B\overline{C}_{\mathcal{W}}+B\overline{C}_{\Gamma_{R_0'}}+B\frac{\log_2(e)}{\epsilon_0}\sqrt{\frac{SN}{\alpha_1'q}}\left(\alpha\left(3\sqrt{2}+1\right)+2(2(\sqrt{2}+1))^{\alpha}\right),
\end{aligned}
\end{equation}
where $B$ is the total bandwidth of the network; $\overline{C}_{\mathcal{W}}$ is the average transport capacity efficiency, defined in terms of meter-bits per second per Hz, of the TX-RX pairs in $\mathcal{W}$; $\overline{C}_{\Gamma_{R_0'}}$ is the average transport capacity efficiency of TX-RX pairs that are not in $\mathcal{W}$ and have communication distances smaller than $R_0'$. The definitions of $\overline{C}_{\mathcal{W}}$ and $\overline{C}_{\Gamma_{R_0'}}$ are formally given below (\ref{eq:g_model_TC_10_TM7}) at the end of Appendix \ref{App:ProofThm7}.

\begin{proof}
This can be proved by following the same procedure as in Appendix \ref{App:ProofThm1}. We thus only illustrate the proof in Appendix \ref{App:ProofThm7}, while some details are omitted for brevity.
\end{proof}

{\em Remark 10:} By using Lemma 3, we conclude that to have a non-vanishing probability for a user to obtain the desired file (i.e., $p_{\text{miss}}(n)$ does not go to $1$), with high probability, the distance between the TX and RX is at least $\Theta\left(\sqrt{\frac{\alpha_1' q}{SN}}\right)$, where $\alpha_1'=\Omega(1)$ and $\alpha_1'=\mathcal{O}\left(q^{\frac{1}{\gamma-1}}\right)$.

\subsection{Outer Bound Analysis for Scenario 1} 

In scenario 1, we assume equal-throughput for users. Then, by following the same procedure as in Sec. \ref{Sec:Out_Bound_Gen}.A and considering Theorem 7 and Remark 10, we can obtain the following theorem:

{\em Theorem 8:} Let $M\to\infty$, $N\to\infty$, and $q\to\infty$. Suppose $\gamma>1$ and $\alpha_1'q=o(M)$. Assume that the powers of users in the network are upper bounded by $P_{\text{max}}$. When $\alpha_1'=\Omega(1)$ is large enough and $\alpha_1'=\mathcal{O}\left(q^{\frac{1}{\gamma-1}}\right)$, the throughput-outage performance of the network is dominated by:
\begin{equation}
\begin{aligned}
T(P_o)&= \Theta\left(\frac{B}{N}\log_2\left(1+\frac{P_{\text{max}}}{N_0B_{\text{s}}}\frac{\chi}{\left(\frac{\alpha_1'q}{SN}\right)^\frac{\alpha}{2}}\right)+B\log_2(e)\left(\alpha\left(3\sqrt{2}+1\right)+2(2(\sqrt{2}+1))^{\alpha}\right)\frac{S}{\alpha_1'q}\right),\\
P_o&= \Theta\left(\frac{1}{(\alpha_1')^{\gamma-1}}\right).
\end{aligned}
\end{equation}
where $P_o$ can be arbitrarily small or converging to zero.
\begin{proof}
The proof can be done by directly using Lemma 3 and Theorem 7 and following the similar proof of proving Theorem 2. We thus omit the proof for brevity.
\end{proof}

{\em Remark 11:} Theorem 8 shows that when $\gamma>1$, $q$ is the dominant factor. In addition, similar to Remark 3, it shows that when the maximum transmit power is some constant, the throughput-outage performance bound considering the generalized physical model has the same scaling law as that considering the protocol model \cite{lee2019throughput}. As a result, this again indicates that the link-level power allocation and scheduling cannot improve the asymptotic throughput-outage performance when users are served under equal-throughput assumption.

\subsection{Outer Bound Analysis for the Proposed Throughput-Enhancing Approach in Scenario 2}

In this subsection, we consider scenario 2, where the network is not confined to the equal-throughput assumption. Similar to the case with $\gamma<1$, we can benefit from letting TX-RX pairs with small communication distances to transmit with a much higher throughput. Hence, we again adopt the throughput-enhancing approach proposed in Sec. \ref{Sec:Out_Bound_Gen}.B, where the first time-slot is used for ordinary file delivery and the second time-slot is used to enhance the overall throughput by letting TX-RX pairs with communication distances smaller than $\sqrt{\epsilon'}R_0'$, where $\epsilon'=\mathcal{O}(1)$, transmit with high speed. By adopting the proposed throughput-enhancing approach, we can obtain the following theorem:

{\em Theorem 9:} Let $M\to\infty$, $N\to\infty$, and $q\to\infty$. Suppose $\gamma>1$ and $\alpha_1'q=o(M)$. Suppose the double time-slot approach discussed in Sec. \ref{Sec:Out_Bound_Gen}.B is used and the $\epsilon'=\mathcal{O}(1)$ is selected. Assume that the powers of users in the network are upper bounded by $P_{\text{max}}$. When $\alpha_1'=\Omega(1)$ is large enough, $\alpha_1'=\mathcal{O}\left(q^{\frac{1}{\gamma-1}}\right)$, and $\lambda_2$ is feasible, the throughput-outage performance of the network is dominated by:
\begin{equation}
\begin{aligned}\label{eq:g_model_TC_Anal_Final_C2_gg1}
T(P_o)=\mathcal{O}\left(\frac{B}{2}\frac{\log_2\left(1+\frac{P_{\text{max}}}{N_0B_{\text{s}}}\frac{\chi}{\left(\frac{\epsilon'\alpha_1'q}{SN}\right)^\frac{\alpha}{2}}\right)}{N}+\frac{B}{2}\frac{S}{\epsilon'\alpha_1'q}\right),P_o= \Theta\left(\frac{1}{(\alpha_1')^{\gamma-1}}\right).
\end{aligned}
\end{equation}
Furthermore, when considering a {\em network instance}, the throughput per user for users with communication distances $d_u>\sqrt{\epsilon'}R_0'$ is dominated by:
\begin{equation}
\lambda_1=\mathcal{O}\left(\frac{B}{N}\log_2\left(1+\frac{P_{\text{max}}}{N_0B_{\text{s}}}\frac{\chi}{\left(\frac{\alpha_1'q}{SN}\right)^\frac{\alpha}{2}}\right)+\frac{BS}{\alpha_1'q}\right);
\end{equation}
the throughput per user for users with communication distances $d_u\leq\sqrt{\epsilon'}R_0'$ is dominated by:
\begin{equation}
\lambda_2=\Theta\left(\frac{B}{2}\frac{\log_2\left(1+\frac{P_{\text{max}}}{N_0B_{\text{s}}}\frac{\chi}{\left(\frac{\epsilon'\alpha_1'q}{SN}\right)^\frac{\alpha}{2}}\right)}{\epsilon'\alpha_1'N}\right)+\Theta\left(\frac{BS}{(\epsilon'\alpha_1')^2q}\right).
\end{equation}

\begin{proof}
See Appendix \ref{App:ProofThm9}.
\end{proof}

By Theorem 9, we understand that minimizing $\epsilon'$ can maximize the overall network throughput. However, there is also a lower bound on $\epsilon'$ due to the physical limitation of the link-rate. The feasibility condition we need to satisfy is again given as:
\begin{equation}\label{eq:lambda2_feas_gg1}
\lambda_2\leq\eta,
\end{equation}
where $\eta$ is some constant. Then, notice that by using Theorem 7, the throughput per user in the second time-slot satisfies
\begin{equation}
\lambda_2\delta' N\Theta\left(\sqrt{\frac{\epsilon'\alpha_1'q}{SN}}\right)=\mathcal{O}\left(B\log_2\left(1+\frac{P_{\text{max}}}{N_0B_{\text{s}}}\frac{\chi}{\left(\frac{\epsilon'\alpha_1'q}{SN}\right)^\frac{\alpha}{2}}\right)\sqrt{\frac{\epsilon'\alpha_1'q}{SN}}+B\sqrt{\frac{SN}{\epsilon'\alpha_1'q}}\right),
\end{equation}
and that $P_{\text{max}}$ is some constant. It follows that we have
\begin{equation}\label{eq:delta_bound_gg1_2}
\lambda_2\delta'=\mathcal{O}\left(\frac{BS}{\epsilon'\alpha_1'q}\right).
\end{equation}
Then, by the derivations in Appendix \ref{App:ProofThm9} and by using the same arguments as in Sec. \ref{Sec:Out_Bound_Gen}.B, we know from (\ref{eq:delta_bound_gg1}) that
\begin{equation}
\delta'=\Theta\left(\epsilon'\alpha_1'\right)
\end{equation}
should be adopted. It follows that we must have
\begin{equation}
\eta(\epsilon'\alpha_1')=\Theta\left(\frac{BS}{\epsilon'\alpha_1'q}\right)
\end{equation}
if we want to minimize $\epsilon'$ subject to the feasibility condition and the tightness of the upper bound. Since $\eta$ is some constant, this then leads to
\begin{equation}\label{eq:epislon_bound_gg1}
\epsilon'\alpha_1'=\Theta\left(\left(\frac{S}{q}\right)^{\frac{1}{2}}\right).
\end{equation}
Finally, by using Theorem 9 and (\ref{eq:epislon_bound_gg1}), we obtain the following theorem:

{\em Theorem 10:} Let $M\to\infty$, $N\to\infty$, and $q\to\infty$. Suppose $\gamma>1$ and $\alpha_1'q=o(M)$. Suppose the double time-slot framework discussed in Sec. \ref{Sec:Out_Bound_Gen}.B is used and the $\epsilon'=\mathcal{O}(1)$ is selected. Assume that the powers of users in the network are upper bounded by $P_{\text{max}}$. When $\alpha_1'=\Omega(1)$ is large enough and $\alpha_1'=\mathcal{O}\left(q^{\frac{1}{\gamma-1}}\right)$, the throughput-outage performance of the network is dominated by:
\begin{equation}
\begin{aligned}
T(P_o)=\Theta\left(\frac{B}{2}\frac{\log_2\left(1+\frac{P_{\text{max}}}{N_0B_{\text{s}}}\frac{\chi}{\left(\left(\frac{q}{S}\right)^{\frac{1}{2}}\frac{1}{N}\right)^\frac{\alpha}{2}}\right)}{N}+\frac{B}{2}\left(\frac{S}{q}\right)^{\frac{1}{2}}\right),P_o=\Theta\left(\frac{1}{(\alpha_1')^{\gamma-1}}\right).
\end{aligned}
\end{equation}
Furthermore, when given a {\em network instance}, the throughput per user for users with communication distances $d_u> \sqrt{\epsilon'}R_0'$ is dominated by:
\begin{equation}
\lambda_1=\Theta\left(\frac{B}{N}\log_2\left(1+\frac{P_{\text{max}}}{N_0B_{\text{s}}}\frac{\chi}{\left(\frac{\alpha_1'q}{SN}\right)^\frac{\alpha}{2}}\right)+\frac{BS}{\alpha_1'q}\right);
\end{equation}
the throughput per user for users with communication distances $d_u\leq\sqrt{\epsilon'}R_0'$ is dominated by:
\begin{equation}
\lambda_2=\Theta\left(\frac{B}{2}\frac{\log_2\left(1+\frac{P_{\text{max}}}{N_0B_{\text{s}}}\frac{\chi}{\left(\left(\frac{q}{S}\right)^{\frac{1}{2}}\frac{1}{N}\right)^\frac{\alpha}{2}}\right)}{\left(\left(\frac{S}{q}\right)^{\frac{1}{2}}\right)N}\right)+\Theta(B).
\end{equation}
\begin{proof}
This is directly obtained by applying (\ref{eq:epislon_bound_gg1}) to Theorem 9.
\end{proof}

{\em Remark 12:} Similar to Remarks 6 and 7, we observe that the double time-slot approach can significantly improve the throughput performance without sacrificing the outage probability. The benefit is again from that we judiciously let TX-RX pairs with very small communication distances transmit at a much higher throughput in the second time-slot. In addition, as indicated by Theorem 10, the double time-slot approach again decouples the tradeoff between throughput and outage and converts the throughput-outage tradeoff to the fairness-outage tradeoff.

\subsection{Achievable Scheme and Analysis for Scenario 1}

Now, we provide the achievable scheme and its corresponding analysis for scenario 1, in which the equal-throughput assumption is considered. We consider the achievable scheme similar to that in Sec. \ref{Sec:Out_Bound_Gen}.C, which is as follows. We first equally split the transmission period $T'$ into two equally-sized time-slots. Then, we let the transmissions in first time-slot follow the TDMA approach and let the transmissions in second time-slot follow the clustering approach. Specifically, for the clustering, we split the network into equally-sized square clusters in which the side length of each cluster is $d=\Theta\left(\sqrt{\frac{\alpha_1 q}{SN}}\right)$. Users in a cluster can only obtain its desired file through users in the same cluster. In each cluster, a user is served at a time and users in the same cluster are served in a round-robin manner. Interference between different clusters are avoided via the frequency reuse approach with reuse factor $(2(K+1))^2$, where $K\in\mathcal{N}>0$ is a finite positive integer. Then, by symmetry of the clusters and by the fact that $N\to\infty$ and that $d=o(1)$, we can follow the same arguments in Sec. \ref{Sec:Out_Bound_Gen}.C such that the number of users in a cluster is given as $N_{\text{cluster}}\backsim F_{\text{Poi}}\left(\frac{\alpha_1 q}{S}\right)$, where $\frac{\alpha_1 q}{S}$ is the mean value of the Poisson distribution. It follows again by Lemma 1 in \cite{lee2020optimal}, the outage probability is 
\begin{equation}
p_o=\sum_{f=1}^M P_r(f) e^{-\frac{\alpha_1 q}{S}P_c(f)},
\end{equation}
where $P_c(f),\forall f$ are determined by the caching policy provided in Theorem 1 of \cite{lee2020optimal}. With the above clustering and caching policy, the same scheduling approach as in Sec. \ref{Sec:Out_Bound_Gen}.C is adopted, and we thus can obtain the following theorem:

{\em Theorem 11:} Let $M\to\infty$, $N\to\infty$, and $q\to\infty$. Suppose that $\gamma>1$ and that $q=o(M)$. Assume that the powers of users in the network are upper bounded by $P_{\text{max}}$. Consider the caching policy in Theorem 1 of \cite{lee2020optimal} and that the side length of a cluster is $\sqrt{\frac{\alpha_1 q}{SN}}$, where $\Theta\left(\frac{\alpha_1 q}{S}\right)=o(M)$. Assume equal-throughput transmissions of users. When $\alpha_1=\Omega(1)$ is large enough and $\alpha_1'=\mathcal{O}\left(q^{\frac{1}{\gamma-1}}\right)$, the following throughput-outage performance of the network is achievable
\begin{equation}
\begin{aligned}
&T(P_o)= \Theta\left(\frac{1}{N}\log_2\left(1+\frac{P_{\text{max}}}{N_0B}\frac{\chi}{\left(\frac{\alpha_1 q}{SN}\right)^\frac{\alpha}{2}}\right)+\frac{BS}{\alpha_1 q}\right),P_o= \Theta\left(\frac{1}{(\alpha_1)^{\gamma-1}}\right),
\end{aligned}
\end{equation}
where $P_o$ can be negligibly small or converging to zero.
\begin{proof}
See Appendix \ref{App:ProofThm11}.
\end{proof}

{\em Remark 13:} By comparing between Theorems 8 and 11, we see that the proposed outer bound is achievable. This indicates that when the equal-throughput assumption is considered, the simple clustering scheme is asymptotically optimal even though we are allowed to use the link-level power allocation and scheduling.

\subsection{Achievable Scheme and Analysis for Scenario 2 with Throughput-Enhancing Approach}

Here, the achievable scheme and the corresponding analysis for scenario 2 are provided. We consider the achievable scheme having the same framework as that in Sec. \ref{Sec:Out_Bound_Gen}.D. Therefore, we again split the transmission duration $T'$ into two time-slots; each has the duration of $\frac{T'}{2}$. We assume without loss of generality that $S$ is an even number. Then, in the first timeslot, we adopt the clustering with side length $d_1=\sqrt{\frac{\alpha_1 q}{SN}}$; in the second timeslot, we adopt the clustering with side length $d_2=\sqrt{\frac{\epsilon\alpha_1 q}{SN}}$, where $\epsilon=\mathcal{O}(1)$. For the caching scheme, we split the whole cache space into two subspace, in which each subspace has size $\frac{S}{2}$. For the first caching subspace, we consider the caching policy proposed in Theorem 1 of \cite{lee2020optimal} with $g_{c,1}(M)=\frac{2\alpha_1 q}{S}$; for the second caching subspace, we adopt the same caching policy and let $g_{c,2}(M)=\frac{2\epsilon\alpha_1 q}{S}$. By the above described scheme, we can then obtain the following theorem:

{\em Theorem 12:} Let $M\to\infty$, $N\to\infty$, and $q\to\infty$. Suppose that $\gamma>1$ and that $q=o(M)$. Assume that the powers of users in the network are upper bounded by $P_{\text{max}}$. Suppose the proposed achievable scheme in Sec. \ref{Sec:Out_Bound_Gen_gg1}.D is used and that the $\epsilon$ is selected such that $\epsilon\alpha_1=\Theta\left(\left(\frac{S}{q}\right)^{\frac{1}{2}}\right)$. Then, when $\alpha_1=\Omega(1)$ is large enough, the following throughput-outage performance of the network is achievable:
\begin{equation}
\begin{aligned}
T(P_o)=\Theta\left(\frac{B}{2}\frac{\log_2\left(1+\frac{P_{\text{max}}}{N_0B_{\text{s}}}\frac{\chi}{\left(\left(\frac{q}{S}\right)^{\frac{1}{2}}\frac{1}{N}\right)^\frac{\alpha}{2}}\right)}{N}+\frac{B}{2}\left(\frac{S}{q}\right)^{\frac{1}{2}}\right),P_o=\Theta\left(\frac{1}{(\alpha_1')^{\gamma-1}}\right).
\end{aligned}
\end{equation}
Furthermore, when given a {\em network instance}, the achievable throughput per user for users with communication distances $d_u\geq \sqrt{\frac{\epsilon'\alpha_1 q}{SN}}$ is:
\begin{equation}
T_1=\Theta\left(\frac{B}{N}\log_2\left(1+\frac{P_{\text{max}}}{N_0B}\frac{\chi}{\left(\frac{\alpha_1 q}{SN}\right)^\frac{\alpha}{2}}\right)+\frac{BS}{\alpha_1 q}\right);
\end{equation}
the achievable throughput per user for users with communication distances $d_u<\sqrt{\frac{\epsilon'\alpha_1 q}{SN}}$ is:
\begin{equation}
T_2=\Theta\left(\frac{B}{2}\frac{\log_2\left(1+\frac{P_{\text{max}}}{N_0B_{\text{s}}}\frac{\chi}{\left(\left(\frac{q}{S}\right)^{\frac{1}{2}}\frac{1}{N}\right)^\frac{\alpha}{2}}\right)}{\left(\left(\frac{S}{q}\right)^{\frac{1}{2}}\right)N}\right)+\Theta(B).
\end{equation}

\begin{proof}
See Appendix \ref{App:ProofThm12}.
\end{proof}

{\em Remark 14:} By comparing between Theorems 10 and 12, we see that the proposed outer bound is achievable. Similar to Remark 9, the achievable scheme of scenario 2 requires the link-level power control and scheduling to enhance the overall throughput. In addition, we stress that although the throughput scaling law $\Theta\left(\sqrt{\frac{S}{q}}\right)$ of scenario 2 gives the same scaling law as the scaling law of the multi-hop cache-aided D2D networks derived in \cite{lee2020optimal}, we should not misinterpret that the single-hop cache-aided D2D with link-level power control and scheduling can have the same performance as the multi-hop cache-aided D2D. This is because comparing our result here with the result derived in \cite{lee2020optimal} is slightly unfair as \cite{lee2020optimal} indeed provides the scaling law considering the equal-throughput assumption. In other word, the single-hop cache-aided D2D with link-level power control and scheduling can have the same performance as its multi-hop counterpart only if the unfairness (from the realization level) is introduced, while the fairness is retained in the multi-hop case.

\section{Throughput-Outage Analysis for Zipf Distribution with $\gamma>1$}

\label{Sec:Zipf_gg1}

In this section, we consider the analysis for networks considering the Zipf distribution, i.e., $q=0$ or equivalently $q=\Theta(1)$. We note that since $M$ is dominant in the case of $\gamma<1$, we are not interested in the case that $\gamma<1$ for the Zipf distribution as the case of Zipf distribution with $\gamma<1$ should have the same throughput-outage scaling law as that of the MZipf distribution with $\gamma<1$. On the other hand, we are interested in the case that $\gamma>1$ for the Zipf distribution only when $P_{\text{max}}\to\infty$ is allowed with $\frac{P_{\text{max}}}{N}=\mathcal{O}(1)$. This is because we already know that when $P_{\text{max}}$ is some constant, the throughput per user is also upper bounded by some constant, and such scaling law has already been achieved without resorting to the link-level power control and scheduling in the literature \cite{lee2019throughput}. Hence, we in this section only focus on whether allowing instantaneous power to be infinity while confining the average power to be constant, i.e., $P_{\text{max}}\to\infty$ with $\frac{P_{\text{max}}}{N}=\mathcal{O}(1)$, can bring additional performance gain. Since we can consider $\gamma>1$ and $q=\Theta(1)$ for deriving the scaling law for the Zipf case, we indeed can use the same procedure of proving Theorems 8 and 11 while considering $q=\Theta(1)$. This then leads to the following theorem:

{\em Theorem 13:} Let $M\to\infty$ and $N\to\infty$. Suppose that $\gamma>1$ and $q=\Theta(1)$. Consider the equal-throughput assumption. When $\alpha_2'=\Theta(1)$ is large enough, the optimal throughput-outage performance of the network is:
\begin{equation}
\begin{aligned}
T(P_o)=\Theta\left(\frac{B}{2}\frac{\log_2\left(1+\frac{P_{\text{max}}}{N_0B_{\text{s}}}\frac{\chi}{\left(\left(\frac{1}{S}\right)^{\frac{\alpha}{2}}\frac{1}{N}\right)^\frac{\alpha}{2}}\right)}{N}\right)+\Theta\left(\frac{BS}{\alpha_2'}\right),P_o=\epsilon_{\text{zip}}(\alpha_2'),
\end{aligned}
\end{equation}
where $\epsilon_{\text{zip}}(\alpha_2')$ can be arbitrarily small. 
\begin{proof}
The proof simply follows the procedure for proving Theorems 8 and 11. We thus omit the proof for brevity.
\end{proof}

By Remark 2, we understand that $\frac{\chi}{\left(\left(\frac{1}{S}\right)^{\frac{\alpha}{2}}\frac{1}{N}\right)^\frac{\alpha}{2}}$ should be upper bounded by $1$. Therefore, from Theorem 13, we observe that the throughput per user is:
\begin{equation}
T(P_o)=\Theta\left(\frac{B}{2}\frac{\log_2\left(1+\frac{P_{\text{max}}}{N_0B_{\text{s}}}\right)}{N}\right)+\Theta\left(\frac{BS}{\alpha_2'}\right).
\end{equation}
Then, since we know $\frac{1}{N}\log_2\left(P_{\text{max}}\right)\to 0$ when $\frac{P_{\text{max}}}{N}=\mathcal{O}(1)$, we obtain the following corollary:

{\em Corollary 2:} Let $M\to\infty$ and $N\to\infty$. Suppose that $\gamma>1$ and $q=\Theta(1)$. Consider the equal-throughput assumption and $\frac{P_{\text{max}}}{N}=\mathcal{O}(1)$. When $\alpha_2'=\Theta(1)$ is large enough, the throughput-outage performance of the network is dominated by:
\begin{equation}
\begin{aligned}
T(P_o)=\Theta\left(\frac{BS}{\alpha_2'}\right),P_o=\epsilon_{\text{zip}}(\alpha_2'),
\end{aligned}
\end{equation}
where $P_o=\epsilon_{\text{zip}}(\alpha_2')$ can be arbitrarily small. 

From Corollary 2, we understand that even if we allow the instantaneous power to be infinity while still confining the average power to be some constant, the best average throughput per user is some constant with the outage probability being arbitrarily small. However, since such throughput-outage performance can be achieved without letting the instantaneous power go to infinity, we conclude the following remark:

{\em Remark 15:} When considering Zipf distribution with $\gamma>1$, even if we allow the instantaneous power to be infinity while still confining to that the constant average power, the throughput-outage performance cannot be improved by the capability of having infinite instantaneous power. Therefore, there is no need to let the instantaneous power to be infinity. In practice, this implies that the network is an interference-limited network, which is in line with our expectation.

{\em Remark 16:} Although the conclusion in Remark 15 is derived assuming the scenario 1, such conclusion applies to scenario 2. This is because it is not possible to obtain an order gain by letting TX-RX pairs with (orderwise) smaller distances to transmit with much higher throughput as the number of TX-RX pairs with (orderwise) smaller distances goes to zero when considering the Zipf distribution with $\gamma>1$. 

\section{Conclusions and Discussions}

This paper investigated the throughput-outage scaling laws of cache-aided single-hop D2D networks considering the generalized physical channel. The main purpose of this paper was to understand whether and how including link-level power control and scheduling can improve the scaling laws. Results showed that when the equal-throughput assumption is considered, i.e., users are served fairly in the realization level, having link-level power control and scheduling cannot improve the performance asymptotically. On the other hand, if the equal-throughput assumption is dropped, i.e., we allow users to have different throughput at the realization level, the scaling laws indeed can be significantly improved by appropriately using link-level power control and scheduling, and then letting TX-RX pairs with small communication distances to transmit at high speed. These results indicated that the approach to benefit from link-level power control and scheduling is to discriminate between TX-RX pairs with different communication distances in the link-level such that the throughput of TX-RX pairs with small communication distances can be significantly enhanced, leading to the asymptotic improvement for the throughput of the network.

Although this paper has conducted the analysis to certain extent, there still remain some possible future directions. First of all, we in this paper focuses only on the double time-slot approach for throughput enhancement in scenario 2, i.e., we only distinguish the TX-RX pairs into two types by their communication distances. Clearly, such approach might be extended to considering more than two types of TX-RX pairs, and whether such extension can bring further performance gain is unclear. Besides, our results are derived under the assumption that users are uniformly distributed within the network. Therefore, our analysis can only be considered as the {\em statistically worst-case} analysis, and different scaling laws might be derived if different user distribution assumptions are considered.

\appendices

\section{Proof of Theorem 1}
\label{App:ProofThm1}

\begin{proof}
Suppose that the delivery are implemented within a period of $T'$ sec The $T'$-second duration is split into different time-slots, in which the duration of a time-slot is $\tau$ sec. We also assume that the total bandwidth $B$ is equally split into $H$ different sub-channels. We denote the bandwidth of a sub-channel as $B_{\text{s}}$ and denote the set of TX-RX pairs that transmit in sub-channel $s$ and time-slot $t$ as $\Gamma_{st}$. Assume $P_{\text{max}}$ is the maximum power that a user can use for transmission; $B_{\text{s}}$. Then, note that a TX-RX pair might be scheduled in more than a time-frequency resource. Then, the total transport capacity for TX-RX pairs in sub-channel $s$ in time-slot $t$, defined in bit-meters, is:
\begin{equation}
\begin{aligned}\label{eq:g_model_subTC}
&C_{\Gamma_{st}}=\sum_{u\in\Gamma_{st}}r_uc_u=\sum_{u\in\Gamma_{st}}r_u\tau B_{\text{s}}\log_2\left(1+\frac{P_ul_{uu^{(\text{r})}}}{N_0B_{\text{s}}+\sum_{k\in\Gamma_{st},k\neq u}P_kl_{ku^{(\text{r})}}}\right)\\
&=r_{w_{st}}\tau B_{\text{s}}\log_2\left(1+\frac{P_{w_{st}}l_{w_{st}w_{st}^{(\text{r})}}}{N_0B_{\text{s}}+\sum_{k\in\Gamma_{st},k\neq w}P_kl_{kw^{(\text{r})}}}\right)+\sum_{u\in\Gamma_{st},u\neq {w_{st}}}r_u\tau B_{\text{s}}\log_2\left(1+\frac{P_ul_{uu^{(\text{r})}}}{N_0B_{\text{s}}+\sum_{k\in\Gamma_{st},k\neq u}P_kl_{ku^{(\text{r})}}}\right)\\
&\leq r_{w_{st}}\tau B_{\text{s}}\log_2\left(1+\frac{P_{w_{st}}l_{w_{st}w_{st}^{(\text{r})}}}{N_0B_{\text{s}}}\right)+\sum_{u\in\Gamma_{st},u\neq {w_{st}}}r_u\tau B_{\text{s}}\log_2\left(1+\frac{P_ul_{uu^{(\text{r})}}}{N_0B_{\text{s}}+\sum_{k\in\Gamma_{st},k\neq u}P_kl_{ku^{(\text{r})}}}\right)\\
& \leq r_{w_{st}}\tau B_{\text{s}}\log_2\left(1+\frac{P_{\text{max}}l_{w_{st}w_{st}^{(\text{r})}}}{N_0B_{\text{s}}}\right)+\sum_{u\in\Gamma_{st},u\neq {w_{st}}}r_u\tau B_{\text{s}}\log_2\left(1+\frac{P_ul_{uu^{(\text{r})}}}{N_0B_{\text{s}}+\sum_{k\in\Gamma_{st},k\neq u}P_kl_{ku^{(\text{r})}}}\right),
\end{aligned}
\end{equation}
where $w_{st}$ is the user who has the largest transmission power in $\Gamma_{st}$, i.e., $P_{w_{st}}\geq P_u,\forall u\in\Gamma_{st}$.

We denote the second term in (\ref{eq:g_model_subTC}) as $C_{\Gamma_{st}}'$ and would like to compute it. To do this, we consider $\Gamma_{st}'=\Gamma_{st}\setminus\lbrace w_{st} \rbrace$ and categorize TX-RX pairs in $\Gamma_{st}'$ into different sets according to their communication distances. Specifically, we let $R_j=\epsilon_0\sqrt{\frac{\rho'M}{SN}}\cdot 2^j,\forall j\in\mathbb{N}$ and define:
\begin{equation}
\begin{aligned}
&\Gamma^{R_0}_{st}=\lbrace u: r_u< R_0,u\in \Gamma_{st}'\rbrace;\\
&\Gamma^{R_j}_{st}=\lbrace u: R_{j-1}\leq r_u< R_j,u\in \Gamma_{st}'\rbrace,\forall j>0.
\end{aligned}
\end{equation}
Then, we can describe $C_{\Gamma_{st}}'$ as
\begin{equation}
\begin{aligned}\label{eq:g_model_subTC_1}
C_{\Gamma_{st}}'&=\tau B_{\text{s}}\left[\sum_{u\in\Gamma^{R_0}_{st}}r_uc_u+\sum_{j>0}\sum_{u\in\Gamma_{st}^{R_j}}r_u\log_2\left(1+\frac{P_ul_{uu^{(\text{r})}}}{N_0B_{\text{s}}+\sum_{k\in\Gamma_{st},k\neq u}P_kl_{ku^{(\text{r})}}}\right)\right]\\
&=\tau B_{\text{s}}\left[\sum_{u\in\Gamma^{R_0}_{st}}r_uc_u+\log_2(e)\sum_{j>0}\sum_{u\in\Gamma_{st}^{R_j}}r_u\log_e\left(1+\frac{P_ul_{uu^{(\text{r})}}}{\sum_{k\in\Gamma_{st},k\neq u}P_kl_{ku^{(\text{r})}}}\right)\right].
\end{aligned}
\end{equation}
We now want to compute 
\begin{equation}
C_{R_j}=\sum_{u\in\Gamma_{st}^{R_j}}r_u\log_e\left(1+\frac{P_ul_{uu^{(\text{r})}}}{\sum_{k\in\Gamma_{st},k\neq u}P_kl_{ku^{(\text{r})}}}\right),\forall j>0.
\end{equation}
To do this, we split the network into equally-sized square clusters whose size length is $R_j$. Then, we collect all clusters that contain at least one RX of the TX-RX pairs in $\Gamma^{R_j}_{st}$ and denote the set of such clusters as $\Phi_j$. Consequently, $\vert\Phi_j\vert\leq \frac{1}{R_j^2}$. We denote the $g$th cluster in $\Phi_j$ as $\phi_g^j$, where $1\leq g\leq \vert\Phi_j\vert$ and denote the number of RXs located in $\phi_g^j$ as $n_g^j$. We denote the communication distance corresponding to RX $h_g^j$ in $\phi_g^j$ as $r_{gh}^j$; denote the transmit power of the TX corresponding to RX $h_g^j$ in $\phi_g^j$ as $P_{gh}^j$, where $h=1,2,...,n_{g}^j$; denote $I_{gh}^j$ as the total interference at RX $h_g^j$ in $\phi_g^j$; and let the indices of the RXs in $\phi_g^j$ follow the descending order of the transmit powers of their corresponding TX, i.e, $P_{gh}^j\geq P_{gh'}^j,\forall h> h'$. Then, we know:
\begin{equation}
\begin{aligned}\label{eq:g_model_subTC_2}
C_{R_j}&=\sum_{u\in\Gamma_{st}^{R_j}}r_u\log_e\left(1+\frac{P_ul_{uu^{(\text{r})}}}{\sum_{k\in\Gamma_{st},k\neq u}P_kl_{ku^{(\text{r})}}}\right)\\
&\leq \underbrace{\sum_{g=1}^{\vert\Phi_j\vert} r_{g1}^j\log_e\left(1+\frac{P_{g1}^j\frac{\chi}{(r_{g1}^j)^{\alpha}}}{I_{g1}^j}\right)}_{(a)}+\underbrace{\sum_{g=1}^{\vert\Phi_j\vert}\sum_{h=2}^{n_g^j} r_{gh}^j\log_e\left(1+\frac{P_{gh}^j\frac{\chi}{(r_{gh}^j)^{\alpha}}}{I_{gh}^j}\right)}_{(b)}
\end{aligned}
\end{equation}
We start computing (\ref{eq:g_model_subTC_2}) by first computing (a) in it. To do this, without loss of generality, we assume $P_{g1}^j\leq P_{g'1}^j$ if $g<g'$. We define $d_g^j=\min\left(\lbrace\sqrt{2}\rbrace\cup\lbrace \vert x_{1_{g'}^j}-x_{1_g^j}\vert:1\leq g'\leq \vert\Phi_j\vert,g'\neq g,g<g' \rbrace\right)$. Then, it should be noted that $w_{st}$, which is the user who has the largest transmit power in $\Gamma_{st}$, must be located within a distance of $\sqrt{2}$ from any receiver $1_{g}^j,\forall g$. Therefore, according to the definition of $d_g^j$, for RX $1_{g}^j$, there must be a TX located within a distance of $d_g^j$ from it. Moreover, such a TX must have a transmit power that is at least as large as $P_{g1}^j$. This thus leads to
\begin{equation}\label{eq:Inter_g1_inequa}
I_{g1}^j\geq P_{g1}^j \frac{\chi}{(d_g^j)^{\alpha}}\geq P_{g1}^j \frac{\chi}{(d_g^j+R_j)^{\alpha}}.
\end{equation}

To proceed, we provide Lemmas 4 and 5 as following:

{\em Lemma 4:} Suppose that the network is split into $\frac{1}{R_j}\times \frac{1}{R_j}$ number of equally-sized square clusters, in which each cluster has size $R_j^2$, where $\frac{1}{R_j}$ is a power of 2. Let $\Delta=\lbrace\delta_1,...,\delta_G\rbrace$ be a set of $G$ points, where $G\leq \frac{1}{R_j^2}$, such that each cluster contains at most one point from $\Delta$. For $g=1,2,...,G,$ we define $d_g^j=\min\left(\lbrace\sqrt{2}\rbrace\cup\lbrace \vert x_{\delta_{g'}}-x_{\delta_g}\vert:1\leq g'\leq G,g'\neq g,g<g' \rbrace\right)$. Then, $\sum_{g=1}^G d_g^j\leq 3\sqrt{2}\frac{1}{R_j}-2\sqrt{2}$.

\begin{proof}
This is obtained by directly applying Lemma 4.2 in \cite{agarwal2004capacity}.
\end{proof}

{\em Lemma 5:} When $x>0$ and $\alpha\geq 1$, we have $\log_e(1+x^{\alpha})\leq \alpha x$.

\begin{proof}
$\log_e(1+x^{\alpha})\leq \log_e\left((1+x)^{\alpha}\right)=\alpha\log_e(1+x)\leq \alpha x$.
\end{proof}

From (\ref{eq:Inter_g1_inequa}), we can upper bound (a) in (\ref{eq:g_model_subTC_2}) as:
\begin{equation}
\begin{aligned}
\sum_{g=1}^{\vert\Phi_j\vert} r_{g1}^j\log_e\left(1+\frac{P_{g1}^j\frac{\chi}{(r_{g1}^j)^{\alpha}}}{I_{g1}^j}\right)\leq \sum_{g=1}^{\vert\Phi_j\vert} r_{g1}^j\log_e\left(1+\frac{P_{g1}^j\frac{\chi}{(r_{g1}^j)^{\alpha}}}{P_{g1}^j\frac{\chi}{(d_g^j+R_j)^{\alpha}}}\right)=\sum_{g=1}^{\vert\Phi_j\vert} r_{g1}^j\log_e\left(1+\frac{(d_g^j+R_j)^{\alpha}}{(r_{g1}^j)^{\alpha}}\right)
\end{aligned}
\end{equation}
Then, by Lemma 5, we obtain:
\begin{equation}
\begin{aligned}
\sum_{g=1}^{\vert\Phi_j\vert} r_{g1}^j\log_e\left(1+\frac{(d_g^j+R_j)^{\alpha}}{(r_{g1}^j)^{\alpha}}\right)\leq \sum_{g=1}^{\vert\Phi_j\vert} r_{g1}^j \alpha\frac{d_g^j+R_j}{r_{g1}^j}=\sum_{g=1}^{\vert\Phi_j\vert} \alpha (d_g^j+R_j).
\end{aligned}
\end{equation}
Finally, by Lemma 4, we obtain:
\begin{equation}
\begin{aligned}
&\sum_{g=1}^{\vert\Phi_j\vert} \alpha (d_g^j+R_j)=\alpha\left(\vert\Phi_j\vert R_j+\sum_{g=1}^{\vert\Phi_j\vert}d_g^j\right)\\
&\stackrel{(a)}{\leq} \alpha\left(\vert\Phi_j\vert R_j+3\sqrt{2}\frac{1}{R_j}-2\sqrt{2}\right)\stackrel{(b)}{\leq}\alpha\left(\frac{1}{R_j}+3\sqrt{2}\frac{1}{R_j}-2\sqrt{2}\right)\leq \frac{\alpha}{R_j}\left(1+3\sqrt{2}\right),
\end{aligned}
\end{equation}
where $(a)$ is due to Lemma 4 and $(b)$ is because $\vert\Phi_j\vert\leq \frac{1}{R_j^2}$. By using the above results, we therefore upper bound (a) of (\ref{eq:g_model_subTC_2}) as:
\begin{equation}\label{eq:g_model_subTC_termA}
\sum_{g=1}^{\vert\Phi_j\vert} r_{g1}^j\log_e\left(1+\frac{P_{g1}^j\frac{\chi}{(r_{g1}^j)^{\alpha}}}{I_{g1}^j}\right)\leq \frac{\alpha}{R_j}\left(3\sqrt{2}+1\right).
\end{equation}

After obtaining the upper bound for (a) of (\ref{eq:g_model_subTC_2}), we now upper bound (b) of (\ref{eq:g_model_subTC_2}). We define
\begin{equation}
P_g^j=\sum_{h=1}^{n_g^j} P_{gh}^j.
\end{equation}
Then, we observe that when we consider only TXs whose RXs are within the same cluster of receiver $u^{(\text{r})}$, the distances between those RXs and RX $u^{(\text{r})}$ must be upper bounded by $\sqrt{2}R_j+R_j$, where the first term is because the largest distance between any receive in the same cluster is $\sqrt{2}R_j$ and the second term is because the the communication distance of the TX-RX pair in $\Gamma_{st}^{R_j}$ is upper bounded by $R_j$. Consequently, the interference $I_{gh}^j$ of a receiver must be lower bounded as:
\begin{equation}
I_{gh}^j\geq (P_g^j-P_{gh}^j)\frac{\chi}{(\sqrt{2}R_j+R_j)^{\alpha}}.
\end{equation}
Furthermore, since we have $P_{g1}^j\geq P_{gh}^j,\forall h\geq 2$, we know $P_g^j\geq 2 P_{gh}^j,\forall h\geq 2$ leading to $P_g^j-P_{gh}^j\geq \frac{P_g^j}{2},\forall h\geq 2$. By the above results, it follows that 
\begin{equation}\label{eq:eq:Inter_gh_inequa}
I_{gh}^j\geq \frac{P_g^j}{2}\frac{\chi}{(\sqrt{2}R_j+R_j)^{\alpha}}.
\end{equation}
By using (\ref{eq:eq:Inter_gh_inequa}), we can then upper bound (b) of (\ref{eq:g_model_subTC_2}) as:
\begin{equation}\label{eq:g_model_subTC_2b}
\sum_{g=1}^{\vert\Phi_j\vert}\sum_{h=2}^{n_g^j} r_{gh}^j\log_e\left(1+\frac{P_{gh}^j\frac{\chi}{(r_{gh}^j)^{\alpha}}}{I_{gh}^j}\right)\leq \sum_{g=1}^{\vert\Phi_j\vert}\sum_{h=2}^{n_g^j} r_{gh}^j\log_e\left(1+\frac{2P_{gh}^j}{P_g^j}\frac{((\sqrt{2}+1)R_j)^{\alpha}}{(r_{gh}^j)^{\alpha}}\right)
\end{equation}
From (\ref{eq:g_model_subTC_2b}), we can further simplify the upper bound. Recall that $R_{j-1}\leq r_{gh}< R_j$ according to the definition of $\Gamma_{st}^{R_j}$. Also, notice that $R_j=2R_{j-1}$ by definition. We can obtain:
\begin{equation}
\begin{aligned}\label{eq:g_model_subTC_termB}
&\sum_{g=1}^{\vert\Phi_j\vert}\sum_{h=2}^{n_g^j} r_{gh}^j\log_e\left(1+\frac{P_{gh}^j}{P_g^j}\frac{((\sqrt{2}+1)R_j)^{\alpha}}{(R^j)^{\alpha}}\right)\leq  \sum_{g=1}^{\vert\Phi_j\vert}\sum_{h=2}^{n_g^j}R^j\log_e\left(1+\frac{2P_{gh}^j}{P_g^j}\frac{(\sqrt{2}+1)^{\alpha}R_j^{\alpha}}{R_{j-1}^{\alpha}}\right)\\
&\leq\sum_{g=1}^{\vert\Phi_j\vert}\sum_{h=2}^{n_g^j} R^j\log_e\left(1+(\sqrt{2}+1)^{\alpha}\frac{2P_{gh}^j}{P_g^j}\frac{2^{\alpha}R_{j-1}^{\alpha}}{R_{j-1}^{\alpha}}\right)=\sum_{g=1}^{\vert\Phi_j\vert}\sum_{h=2}^{n_g^j} R^j\log_e\left(1+(2(\sqrt{2}+1))^{\alpha}\frac{2P_{gh}^j}{P_g^j}\right)\\
&\stackrel{(a)}{\leq}\sum_{g=1}^{\vert\Phi_j\vert}\sum_{h=2}^{n_g^j} R^j(2(\sqrt{2}+1))^{\alpha}\frac{2P_{gh}^j}{P_g^j}=2R^j(2(\sqrt{2}+1))^{\alpha}\sum_{g=1}^{\vert\Phi_j\vert}\sum_{h=2}^{n_g^j} \frac{P_{gh}^j}{P_g^j}= 2R^j(2(\sqrt{2}+1))^{\alpha}\sum_{g=1}^{\vert\Phi_j\vert}\frac{\sum_{h=2}^{n_g^j}P_{gh}^j}{P_g^j}\\
&\leq 2R^j(2(\sqrt{2}+1))^{\alpha}\sum_{g=1}^{\vert\Phi_j\vert}\frac{\sum_{h=1}^{n_g^j}P_{gh}^j}{P_g^j}\stackrel{(b)}{=}2R^j(2(\sqrt{2}+1))^{\alpha}\sum_{g=1}^{\vert\Phi_j\vert}1=2(2(\sqrt{2}+1))^{\alpha}R^j\vert\Phi_j\vert \\
&\stackrel{(c)}{\leq}2(2(\sqrt{2}+1))^{\alpha}R^j\frac{1}{R_j^2}=2(2(\sqrt{2}+1))^{\alpha}\frac{1}{R_j},
\end{aligned}
\end{equation}
where $(a)$ is because $\log_e(1+x)\leq x$ when $x>0$; $(b)$ is because $\sum_{h=1}^{n_g^j}P_{gh}^j=P_g^j$ by definition; and $(c)$ is because $\vert\Phi_j\vert\leq\frac{1}{R_j^2}$.

We note that results in (\ref{eq:g_model_subTC_termA}) and (\ref{eq:g_model_subTC_termB}) can be applied when $j>0,\forall j$. Then, by using (\ref{eq:g_model_subTC_2}), (\ref{eq:g_model_subTC_termA}), and (\ref{eq:g_model_subTC_termB}), we can further upper bound the second term in (\ref{eq:g_model_subTC_1}) as follows:
\begin{equation}
\begin{aligned}\label{eq:g_model_subTC_3}
&\sum_{j>0}\sum_{u\in\Gamma_{st}^{R_j}}r_u\log_e\left(1+\frac{P_ul_{uu^{\text{(r)}}}}{\sum_{k\in\Gamma_{st},k\neq u}P_kl_{ku^{\text{(r)}}}}\right)\leq\sum_{j>0}\left(\left(3\sqrt{2}+1\right)\frac{\alpha}{R_j}+2(2(\sqrt{2}+1))^{\alpha}\frac{1}{R_j}\right)\\
&=\left(\alpha\left(3\sqrt{2}+1\right)+2(2(\sqrt{2}+1))^{\alpha}\right)\sum_{j>0}\frac{1}{R_j}\stackrel{(a)}{=}\frac{\left(\alpha\left(3\sqrt{2}+1\right)+2(2(\sqrt{2}+1))^{\alpha}\right)}{R_0}\sum_{j=1}^{\infty}2^{-j}\\
&=\frac{\left(\alpha\left(3\sqrt{2}+1\right)+2(2(\sqrt{2}+1))^{\alpha}\right)}{R_0}=\left(\alpha\left(3\sqrt{2}+1\right)+2(2(\sqrt{2}+1))^{\alpha}\right)\frac{1}{\epsilon_0}\sqrt{\frac{SN}{\rho'M}},
\end{aligned}
\end{equation}
where $(a)$ is because $R_j=2R_{j-1}$. By combining (\ref{eq:g_model_subTC}), (\ref{eq:g_model_subTC_1}), and (\ref{eq:g_model_subTC_3}), we obtain
\begin{equation}
\begin{aligned}\label{eq:g_model_subTC_final}
C_{\Gamma_{st}}\leq& r_{w_{st}}\tau B_{\text{s}}\log_2\left(1+\frac{P_{\text{max}}l_{w_{st}w_{st}^{\text{(r)}}}}{N_0B_{\text{s}}}\right)+\tau B_{\text{s}}\sum_{u\in\Gamma^{R_0}_{st}}r_uc_u\\
&+\tau B_{\text{s}}\frac{\log_2(e)}{\epsilon_0}\sqrt{\frac{SN}{\rho'M}}\left(\alpha\left(3\sqrt{2}+1\right)+2(2(\sqrt{2}+1))^{\alpha}\right)
\end{aligned}
\end{equation}
Finally, by using (\ref{eq:g_model_subTC_final}) and summing contributions in all time-slots and sub-channels, we obtain
\begin{equation}
\begin{aligned}\label{eq:g_model_TC}
&C_{\Gamma}=\frac{1}{T'}\sum_{t}\sum_{s}C_{\Gamma_{st}}\leq \frac{1}{T'}\sum_{t}\sum_{s}\tau B_{\text{s}}\\
&\cdot\left(r_{w_{st}}\log_2\left(1+\frac{P_{\text{max}}l_{w_{st}w_{st}^{\text{(r)}}}}{N_0B_{\text{s}}}\right)+\sum_{u\in\Gamma^{R_0}_{st}}r_uc_u+\frac{\log_2(e)}{\epsilon_0}\sqrt{\frac{SN}{\rho'M}}\left(\alpha\left(3\sqrt{2}+1\right)+2(2(\sqrt{2}+1))^{\alpha}\right)\right)\\
&=\frac{\tau B_{\text{s}}}{T'}\sum_{t}\sum_{s}r_{w_{st}}\log_2\left(1+\frac{P_{\text{max}}l_{w_{st}w_{st}^{\text{(r)}}}}{N_0B_{\text{s}}}\right)+\frac{\tau B_{\text{s}}\sum_{t}\sum_{s}\sum_{u\in\Gamma^{R_0}_{st}}r_uc_u}{T'}\\
&\qquad\qquad+B\frac{\log_2(e)}{\epsilon_0}\sqrt{\frac{SN}{\rho'M}}\left(\alpha\left(3\sqrt{2}+1\right)+2(2(\sqrt{2}+1))^{\alpha}\right)\\&
=B\overline{C}_{\mathcal{W}}+B\overline{C}_{\Gamma_{R_0}}+B\frac{\log_2(e)}{\epsilon_0}\sqrt{\frac{SN}{\rho'M}}\left(\alpha\left(3\sqrt{2}+1\right)+2(2(\sqrt{2}+1))^{\alpha}\right),
\end{aligned}
\end{equation}
where $\mathcal{W}=\lbrace w_{st},\forall s,t\rbrace$,
\begin{equation}
\overline{C}_{\mathcal{W}}=\frac{\tau B_{\text{s}}\sum_{t}\sum_{s}r_{w_{st}}\log_2\left(1+\frac{P_{\text{max}}l_{w_{st}w_{st}^{\text{(r)}}}}{N_0B_{\text{s}}}\right)}{BT'}
\end{equation}
is the average transport capacity, defined in terms of per Hz per second, of the transmitter-receiver pairs whose powers are the largest among their corresponding time-frequency resource and 
\begin{equation}
\overline{C}_{\Gamma_{R_0}}=\frac{\tau B_{\text{s}}\sum_{t}\sum_{s}\sum_{u\in\Gamma^{R_0}_{st}}r_uc_u}{BT'}
\end{equation}
is the average transport capacity of the TX-RX pairs that do not have the largest powers among their corresponding time-frequency resource and have communication distance smaller than $R_0$.

\end{proof} 

\section{Proof of Theorem 2}

\label{App:ProofThm2}

To have a negligibly small outage probability, according to Remark 1, we need TX-RX pairs to have communication distances $r_u=\Theta\left(\sqrt{\frac{\rho'M}{SN}}\right)$ with high probability, where $\rho'=\Omega(1)$. This indicates that the average distance of a bit that is transported in the network is $\overline{L}=\Theta\left(\sqrt{\frac{\rho'M}{SN}}\right)$. From (\ref{eq:g_model_TC_Thm}), we first observe that the upper bound can be optimized by maximizing $\overline{C}_{\mathcal{W}}$ and $\overline{C}_{\Gamma_{R_0}}$. When optimizing $\overline{C}_{\mathcal{W}}$, we aim to have larger $r_w,\forall w\in \mathcal{W}$. In addition, we know that, w.h.p., the communication
distances of users are $r_u=\mathcal{O}\left(\sqrt{\frac{\rho'M}{SN}}\right),\forall u\in\Gamma$. Hence, we can assume without loss of optimality that $\overline{r}_{w}=\Theta\left(\sqrt{\frac{\rho'M}{SN}}\right),\forall w\in \mathcal{W}$. Therefore, we obtain:
\begin{equation}
\begin{aligned}\label{eq:g_model_TC_Anal_1}
C_{\Gamma}&\leq B\Theta\left(\sqrt{\frac{\rho'M}{SN}}\log_2\left(1+\frac{P_{\text{max}}}{N_0B_{\text{s}}}\frac{\chi}{\left(\frac{\rho'M}{SN}\right)^\frac{\alpha}{2}}\right)\right)\\
&+B\overline{C}_{\Gamma_{R_0}}+B\frac{\log_2(e)}{\epsilon_0}\sqrt{\frac{SN}{\rho'M}}\left(\alpha\left(3\sqrt{2}+1\right)+2(2(\sqrt{2}+1))^{\alpha}\right),
\end{aligned}
\end{equation} 
We then denote $\lambda$ as the average throughput per user (bits/s) in $T$ sec; let $\Gamma^{R_0}_{\text{tot}}=\lbrace u: \vert x_u-x_{u^{\text{r}}}\vert<R_0,u\in\Gamma\setminus \mathcal{W} \rbrace$; and let 
\begin{equation}
C_{P_{\text{max}}}= B\Theta\left(\sqrt{\frac{\rho'M}{SN}}\log_2\left(1+\frac{P_{\text{max}}}{N_0B_{\text{s}}}\frac{\chi}{\left(\frac{\rho'M}{SN}\right)^\frac{\alpha}{2}}\right)\right).
\end{equation}
By definition, we obtain
\begin{equation}
\begin{aligned}
&C_{\Gamma}=\lambda N \overline{L}=\lambda N\Theta\left(\sqrt{\frac{\rho'M}{SN}}\right)=\lambda\Theta\left(\sqrt{\frac{\rho'MN}{S}}\right)\\
&\leq C_{P_{\text{max}}}+ B\overline{C}_{\Gamma_{R_0}}+B\frac{\log_2(e)}{\epsilon_0}\sqrt{\frac{SN}{\rho'M}}\left(\alpha\left(3\sqrt{2}+1\right)+2(2(\sqrt{2}+1))^{\alpha}\right)\\
&\stackrel{(a)}{\leq} C_{P_{\text{max}}}+ \sum_{u\in\Gamma^{R_0}_{\text{tot}}}\epsilon_0\sqrt{\frac{\rho'M}{SN}}\lambda+B\frac{\log_2(e)}{\epsilon_0}\sqrt{\frac{SN}{\rho'M}}\left(\alpha\left(3\sqrt{2}+1\right)+2(2(\sqrt{2}+1))^{\alpha}\right)
\end{aligned}
\end{equation}
where $(a)$ is because $r_u<R_0,\forall u\in\Gamma^{R_0}_{\text{tot}}$ by definition and because we adopt the equal-throughput assumption. It follows that 
\begin{equation}
\begin{aligned}
\lambda\leq C_{P_{\text{max}}}\sqrt{\frac{S}{\rho'MN}}+\lambda\epsilon_0\frac{1}{N}\sum_{u\in\Gamma^{R_0}_{\text{tot}}}1+B\frac{\log_2(e)}{\epsilon_0}\frac{S}{\rho'M}\left(\alpha\left(3\sqrt{2}+1\right)+2(2(\sqrt{2}+1))^{\alpha}\right).
\end{aligned}
\end{equation}
We let the number of users having communication distances smaller than $R_0$ be $\epsilon N$, where $\epsilon=\mathcal{O}(1)<1$. This leads to
\begin{equation}
\lambda(1-\epsilon_0\epsilon)\leq B\Theta\left(\frac{1}{N}\log_2\left(1+\frac{P_{\text{max}}}{N_0B_{\text{s}}}\frac{\chi}{\left(\frac{\rho'M}{SN}\right)^\frac{\alpha}{2}}\right)\right)+B\frac{\log_2(e)}{\epsilon_0}\frac{S}{\rho'M}\left(\alpha\left(3\sqrt{2}+1\right)+2(2(\sqrt{2}+1))^{\alpha}\right).
\end{equation}
It follows that 
\begin{equation}\label{eq:g_model_TC_Anal_Final_C1}
\lambda\leq \frac{B}{(1-\epsilon_0\epsilon)}\left(\Theta\left(\frac{1}{N}\log_2\left(1+\frac{P_{\text{max}}}{N_0B_{\text{s}}}\frac{\chi}{\left(\frac{\rho'M}{SN}\right)^\frac{\alpha}{2}}\right)\right)+\frac{\log_2(e)\left(\alpha\left(3\sqrt{2}+1\right)+2(2(\sqrt{2}+1))^{\alpha}\right)}{\epsilon_0}\frac{S}{\rho'M}\right).
\end{equation}
Recall that, w.h.p., the communication distances of TX-RX pairs is $\Theta\left(\sqrt{\frac{\rho'M}{SN}}\right)$ and $\epsilon$ is small when $\epsilon_0$ is small. Thus, $(1-\epsilon_0\epsilon)$ should be some constant. Then, from (\ref{eq:g_model_TC_Anal_Final_C1}) and Lemma 2, we complete the proof.

\section{Proof of Theorem 3}

\label{App:ProofThm3}

With the described framework in Sec. \ref{Sec:Out_Bound_Gen}.B, the throughput per user in the first time-slot is: 
\begin{equation}
\lambda_1=\mathcal{O}\left(\frac{B}{N}\log_2\left(1+\frac{P_{\text{max}}}{N_0B_{\text{s}}}\frac{\chi}{\left(\frac{\rho'M}{SN}\right)^\frac{\alpha}{2}}\right)+\frac{BS}{\rho'M}\right).
\end{equation}
Subsequently, since we delivery files only for those TX-RX pairs whose communication distances are smaller than $d_u=\sqrt{\epsilon'} R_0$, where $\epsilon'=\mathcal{O}(1)$, by using the same analysis as for scenario 1, we can obtain the throughput per user in the second time-slot satisfies:
\begin{equation}\label{eq:out_lambda_2}
\lambda_2\delta' N\Theta\left(\sqrt{\frac{\epsilon'\rho'M}{SN}}\right)=\mathcal{O}\left(B\log_2\left(1+\frac{P_{\text{max}}}{N_0B_{\text{s}}}\frac{\chi}{\left(\frac{\epsilon'\rho'M}{SN}\right)^\frac{\alpha}{2}}\right)\sqrt{\frac{\epsilon'\rho'M}{SN}}+B\sqrt{\frac{SN}{\epsilon'\rho'M}}\right),
\end{equation}
where $\lambda_2$ is the throughput per user for users transmitting in the second time-slot and $\delta'N$ is the number of users that have $d_u\leq\sqrt{\epsilon'} R_0$. Note that $\delta'$ is a function of $\epsilon'$ and $R_0$. By combining the results in both time-slots, it follows that the overall throughput per user is
\begin{equation}
\begin{aligned}\label{eq:out_lambda_final}
&\lambda=\mathcal{O}\left(\frac{\lambda_1 N+ \lambda_2\delta' N }{2N}\right)\\
&=\mathcal{O}\left(\frac{B}{2}\frac{\log_2\left(1+\frac{P_{\text{max}}}{N_0B_{\text{s}}}\frac{\chi}{\left(\frac{\rho'M}{SN}\right)^\frac{\alpha}{2}}\right)}{N}+\frac{B}{2}\frac{S}{\rho'M}\right)+\mathcal{O}\left(\frac{B}{2}\frac{\log_2\left(1+\frac{P_{\text{max}}}{N_0B_{\text{s}}}\frac{\chi}{\left(\frac{\epsilon'\rho'M}{SN}\right)^\frac{\alpha}{2}}\right)}{N}+\frac{B}{2}\frac{S}{\epsilon'\rho'M}\right),
\end{aligned}
\end{equation}
where the outage lower bound is $P_o=\Theta\left(e^{-\rho'}\right)$. Since $\epsilon'=\mathcal{O}(1)$, we notice that in this case $\lambda$ is dominated by the second term, implying that the overall network throughput is enhanced by the throughput generated in the second time-slot, while the outage probability is maintained low by the transmissions in the first time-slot. The above discussions then lead to Theorem 3.

\section{Proof of Corollary 1}

\label{App:ProofCoro1}

Observe that $\delta'$ can make an impact on $\lambda_2$, and $\lambda_2$ can characterize the upper bound of the throughput-enhancing ability. Hence, we need to characterize $\delta'$ to better characterize $\lambda_2$. To do this, we need to obtain the maximal probability that users can obtain their desired files from users within the distance of $\sqrt{\epsilon'} R_0$. This is equivalent to finding the minimal outage probability within a cluster whose side length is $\Theta\left(\sqrt{\epsilon'} R_0\right)$. Then, by Proposition 2 of \cite{lee2020optimal}, we know that the minimal outage probability in this case is
\begin{equation}
\begin{aligned}\label{eq:Pout_sg}
p_o^{\text{sec}}= &1+(1-\gamma)e^{-\gamma\left(\frac{1}{C_1}-1\right)}\left(\frac{C_1S}{\gamma}\frac{g_c'}{M}\right)^{1-\gamma}\frac{\left(\frac{C_1}{C_1+C_2}\right)^{\gamma}\cdot\left(\frac{C_2}{C_1+C_2}\right)^{\gamma\frac{ C_2}{C_1}}}{\left(1+\frac{C_2S}{\gamma}\frac{g_c(M)}{M}\right)^{1-\gamma}-\left(\frac{C_2S}{\gamma}\frac{g_c(M)}{M}\right)^{1-\gamma}}\\
&\qquad - \left(\frac{C_1S}{\gamma}\frac{g_c(M)}{M}\right)^{1-\gamma}\frac{\left(1+\frac{C_2}{C_1}\right)^{1-\gamma}-\left(\frac{C_2}{C_1}\right)^{1-\gamma}}{\left(1+\frac{C_2S}{\gamma}\frac{g_c(M)}{M}\right)^{1-\gamma}-\left(\frac{C_2S}{\gamma}\frac{g_c(M)}{M}\right)^{1-\gamma}},
\end{aligned}
\end{equation} 
where $g_c'=\Theta\left(\epsilon'R_0^2N\right)=\Theta\left(\epsilon'\rho'\frac{M}{S}\right)$, $C_2=\frac{q\gamma}{Sg_c'}$, and $C_1$ is the solution of the equation: $C_1=1+C_2\log\left(1+\frac{C_1}{C_2}\right)$. Then, we assume $q=\mathcal{O}\left(\epsilon'\rho'\frac{M}{S}\right)$. It follows that $C_1=\Theta(1)$ and $C_2=\mathcal{O}(1)$. Hence, (\ref{eq:Pout_sg}) can be written as
\begin{equation}
p_o^{\text{sec}}=1+\Theta\left((1-\gamma)e^{\gamma}\left(\frac{S}{\gamma}\frac{g_c'}{M}\right)^{1-\gamma}\right)-\Theta\left(\left(\frac{S}{\gamma}\frac{g_c'}{M}\right)^{1-\gamma}\right)
\end{equation}
It follows that maximal probability for a user to find its desired file from users within the  distance of $\sqrt{\epsilon'}R_0$ is:
\begin{equation}\label{eq:Phit_2_SecIIIA}
p_{\text{h}}^{\text{sec}}=\Theta\left(\left(\frac{S}{\gamma}\frac{g_c'}{M}\right)^{1-\gamma}\right)=\Theta((\epsilon'\rho')^{1-\gamma}).
\end{equation}
This leads to
\begin{equation}\label{eq:delta_bound_SecIIIA}
\delta'=\Theta\left((\epsilon'\rho')^{1-\gamma}\right).
\end{equation}
Combining Theorem 3 and (\ref{eq:delta_bound_SecIIIA}), we complete the proof.

\section{Proof of Proposition 1}
\label{App:ProofProp1}
\begin{proof}
By the using frequency reuse scheme with reuse factor $(2(K+1))^2$ to avoid the inter-cluster interference, interference for a TX in a specifc cluster comes from users in other clusters that transmit at the same frequency. We then notice that (see Fig. 4 in \cite{franceschetti2007closing}) the closest TXs causing interference is located in a square at distance at least $K+1$ clusters away. Specifically, the closet ring of the $8$ interfering TXs are $K+1$ clusters away; the second closest ring of the $16$ interfering TXs are at least $(3K+3)$ clusters away. Overall, the upper bound of the total interference power can be summarized as following:
\begin{equation}
\begin{aligned}\label{eq:Interference_BPHY}
I(d,K)&\leq \sum_{i=1}^{\infty} 8i\nu_{\text{upp}}\frac{\chi}{\left(d(2i-1)(K+1)\right)^{\alpha}}= \frac{8\nu_{\text{upp}}\chi}{\left(d(K+1)\right)^{\alpha}}\sum_{i=1}^{\infty}\frac{i}{(2i-1)^{\alpha}}\\
&\leq \frac{8\nu_{\text{upp}}\chi}{\left(d(K+1)\right)^{\alpha}}\sum_{i=1}^{\infty}\frac{i}{i^{\alpha}}=\frac{8\nu_{\text{upp}}\chi}{\left(d(K+1)\right)^{\alpha}}\sum_{i=1}^{\infty}i^{1-\alpha}.
\end{aligned}
\end{equation}
We denote $\sum_{i=1}^{\infty}i^{1-\alpha}$ as $I_{\text{c}}$, which clearly converges when $\alpha>2$. Then, since the distance between the TX and the RX in the same cluster is at most $\sqrt{2}d$, we know that 
\begin{equation}\label{eq:Signal_BPHY}
S(d)\geq \frac{\nu_{\text{low}}\chi}{(\sqrt{2}d)^{\alpha}}.
\end{equation}
By observing (\ref{eq:Interference_BPHY}) and (\ref{eq:Signal_BPHY}), we then conclude that the SINR of a D2D link in a cluster is:
\begin{equation}\label{eq:SINR_BPHY_1}
\text{SINR}_{\text{D2D}}\geq \frac{\frac{\nu_{\text{low}}\chi}{(\sqrt{2}d)^{\alpha}}}{B_u N_0+\frac{8\nu_{\text{upp}}\chi}{\left(d(K+1)\right)^{\alpha}}I_{\text{c}}}.
\end{equation}
Then, observe that 
\begin{equation}\label{eq:SINR_BPHY_2}
\frac{\frac{\nu_{\text{low}}\chi}{(\sqrt{2}d)^{\alpha}}}{\frac{8\nu_{\text{upp}}\chi}{\left(d(K+1)\right)^{\alpha}}I_{\text{c}}}=\frac{\nu_{\text{low}}}{8\nu_{\text{upp}}I_{\text{c}}}\left(\frac{K+1}{\sqrt{2}}\right)^{\alpha}=\Theta(1)
\end{equation}
is lower bounded by some constant $\vartheta'$ and is monotonically increasing with respect to $K$. Furthermore, since the minimum transmit power is some constant, there must exist some constant $\vartheta''$ such that $\frac{\nu_{\text{low}}\chi}{B_uN_0(\sqrt{2}d)^{\alpha}}\geq \vartheta''$. It follows that for any activated TX-RX pair $u$ in each cluster, there must exist some constant $\vartheta$ such that the rate of the TX-RX pair is
\begin{equation}
R(u,u^{(\text{r})})=B_u\log_2\left(1+\frac{P_ul_{uu^{(\text{r})}}}{B_uN_0+\sum_{k\neq u,k\in\Gamma^u_{\text{Co}}}P_kl_{ku^{(\text{r})}}}\right)\geq B_u\log_2(1+\vartheta),
\end{equation}
where $B_u=\frac{B}{(2(K+1))^2}$. Note that $\vartheta$ is monotonically increasing with respect to the reuse factor $K$ as (\ref{eq:SINR_BPHY_2}) is monotonically increasing with respect to $K$. This completes the proof.
\end{proof}

\section{Proof of Theorem 5}
\label{App:ProofThm5}
\begin{proof}

We first derive the outage probability. Note that the number of users $N_{\text{cluster}}$ in a cluster follows the Poisson distribution whose mean value is $\frac{\rho M}{S}$. Then, according to Proposition 1 of \cite{lee2020optimal}, when $\rho$ is sufficiently large, the outage probability is upper bounded as
\begin{equation}
p_o\leq\Theta\left(e^{-\rho}\right),
\end{equation}
indicating that the outage probability can be negligibly small or converging to zero. Furthermore, since the mean number of users in a cluster is $\frac{\rho M}{S}\to \infty$ and $p_0$ is either negligibly small or converging to zero, we can conclude that for each cluster, the probability that all users are in outage goes to zero; thus the probability that a cluster can have at least a TX-RX pair to schedule goes to 1.

Now, we derive the network throughput. We first note that due to symmetry of the clusters and due to that users in a cluster are served in a round-robin manner, the proposed achievable scheme satisfies the equal-throughput assumption. Then, for the first time-slot, since the simple TDMA is used for users, the achievable network throughput of the first time-slot is:
\begin{equation}\label{eq:Ach_Tnet1}
T_{\text{net},1}=B\log_2\left(1+\frac{P_{\text{max}}}{N_0B}\frac{\chi}{\left(\frac{\rho M}{SN}\right)^\frac{\alpha}{2}}\right)
\end{equation}
Subsequently, recall that by Proposition 1 in Sec. \ref{Sec:Out_Bound_Gen}.C, any TX-RX pair in a cluster can be activated with an appropriate $K$. Thus, for the second sub-timeslot, since the number of clusters in the network is at least $\left\lfloor\frac{SN}{\rho M}\right\rfloor$, by symmetry of clusters, the network throughput can be obtained as follows:
\begin{equation}
T_{\text{net},2}\geq \mathbb{E}\left[\sum_{c=1}^{\left\lfloor\frac{SN}{\rho M}\right\rfloor}T_{c}\right]=\left\lfloor\frac{SN}{\rho M}\right\rfloor\cdot\mathbb{E}\left[T_{c}\right],
\end{equation}
where $T_{\text{c}}$ is the link-rate of cluster $c$ in the second time-slot. Since the probability that a cluster can have at least a TX-RX link to schedule goes to 1 and any scheduled TX-RX link can provide at least the the rate $\frac{B}{(2(K+1))^2}\log_2\left(1+\vartheta\right)$ according to Proposition 1, it follows that the achievable network throughput generated at the second time-slot is
\begin{equation}\label{eq:Ach_Tnet2}
T_{\text{net},2}\geq\left\lfloor\frac{SN}{\rho M}\right\rfloor\cdot\frac{B}{(2(K+1))^2}\log_2\left(1+\vartheta\right).
\end{equation}
Due to symmetry of users and by using (\ref{eq:Ach_Tnet1}) and (\ref{eq:Ach_Tnet2}), the overall achievable throughput per user is
\begin{equation}
T(P_o)=\frac{T_{\text{net},1}+T_{\text{net},2}}{2N}=\underbrace{\Theta\left(\frac{B}{N}\log_2\left(1+\frac{P_{\text{max}}}{N_0B}\frac{\chi}{\left(\frac{\rho M}{SN}\right)^\frac{\alpha}{2}}\right)\right)}_{\text{come from the first timeslot}}+\underbrace{\Theta\left(\frac{BS}{\rho M}\right)}_{\text{come from the second timeslot}},
\end{equation}
where the corresponding achievable outage probability is 
\begin{equation}
P_o= \Theta\left(e^{-\rho}\right).
\end{equation}
\end{proof}

\section{Proof of Theorem 6}

\label{App:ProofThm6}

Recall that we split the whole cache space into two subspaces, and each has size $\frac{S}{2}$. For the first caching subspace, we consider the cluster size $d_1=\sqrt{\frac{\rho M}{SN}}$ and adopt the caching policy proposed in Theorem 1 of \cite{lee2020optimal} with $g_{c,1}(M)=\frac{2\rho M}{S}$. Thus, by using Theorem 5, the achievable throughput-outage performance for users receiving files in the first time-slot is given as:
\begin{equation}
\begin{aligned}
&T_1(P_o)= \Theta\left(\frac{1}{N}\log_2\left(1+\frac{P_{\text{max}}}{N_0B}\frac{\chi}{\left(\frac{2\rho M}{SN}\right)^\frac{\alpha}{2}}\right)+\frac{BS}{2\rho M}\right),P_o= \Theta\left(e^{-\rho}\right).
\end{aligned}
\end{equation}
Then, for the second caching subspace, we adopt the caching policy proposed in Theorem 1 of \cite{lee2020optimal} and let $g_{c,2}(M)=\frac{2\epsilon\rho M}{S}$. Note that since we let $d_2=\sqrt{\frac{2\epsilon\rho M}{SN}}$ in the second time-slot, the overall probability that a user can find their desired in the cluster is lower bounded by the corresponding probability that only the second caching subspace is considered. Then, according to Proposition 2 of \cite{lee2020optimal} and derivations in Sec. \ref{Sec:Out_Bound_Gen}.B, when $q=\mathcal{O}\left(\epsilon\rho \frac{M}{S}\right)$, such probability is given by (\ref{eq:Phit_2_SecIIIA}), where
\begin{equation}
p_{\text{h}}^{\text{sec}}\geq\Theta((\epsilon\rho)^{1-\gamma}).
\end{equation}
Then, supposing
\begin{equation}\label{eq:Ach_re}
\epsilon\rho=\Theta\left(\left(\frac{S}{M}\right)^{\frac{1}{2-\gamma}}\right),
\end{equation}
the average number of users in a cluster that can find the desired files in the second time-slot is lower bounded by
\begin{equation}
\Theta\left(\frac{2\epsilon\rho M}{S}p_{\text{h}}^{\text{sec}}\right)=\Theta\left(1\right),
\end{equation}
which can be a large constant if we let $\epsilon\rho$ in (\ref{eq:Ach_re}) to have a large constant factor, namely, $\epsilon\rho=C_{\text{sec}}\left(\frac{S}{M}\right)^{\frac{1}{2-\gamma}}$, where $C_{\text{sec}}$ is a large constant. It follows that the probability of activating a cluster with side length $d_2=\sqrt{\frac{2\epsilon\rho M}{SN}}$ can be close to 1. Then, note that we have 
\begin{equation}
\frac{1}{d_2^2}=\frac{N S}{2\rho\epsilon M}
\end{equation}
number of such cluster. Besides, by Proposition 1, we know the link-rate can be $B\log_2(1+\vartheta)$, where $\vartheta$ is some constant. It follows that the network throughput in the second time-slot can be given as
\begin{equation}
\Theta\left(\frac{BN S}{2\rho\epsilon M}\right).
\end{equation}
Since the number of users in the network that can receive files in the second time-slot is
\begin{equation}
Np_{\text{h}}^{\text{sec}}=\Theta\left(N(\epsilon\rho)^{1-\gamma}\right),
\end{equation}
the throughput per user for users that can receive files in the second time-slot is
\begin{equation}
\begin{aligned}
T_2&=\Theta\left(\frac{B}{2}\frac{\log_2\left(1+\frac{P_{\text{max}}}{N_0B_{\text{s}}}\frac{\chi}{\left(\frac{\epsilon\rho M}{SN}\right)^\frac{\alpha}{2}}\right)}{(\epsilon\rho)^{1-\gamma}N}\right)+\Theta\left(\frac{BN S}{2\rho\epsilon M}\frac{1}{N(\epsilon\rho)^{1-\gamma}}\right)\\
&=\Theta\left(\frac{B}{2}\frac{\log_2\left(1+\frac{P_{\text{max}}}{N_0B_{\text{s}}}\frac{\chi}{\left(\frac{\epsilon\rho M}{SN}\right)^\frac{\alpha}{2}}\right)}{(\epsilon\rho)^{1-\gamma}N}\right)+\Theta\left(\frac{BS}{M(\epsilon\rho)^{2-\gamma}}\right).
\end{aligned}
\end{equation}
By combining results in the first and second time-slots and by symmetry of users, we can see that the average throughput per users is:
\begin{equation}
T(P_o)=\Theta\left(\frac{B}{2}\frac{\log_2\left(1+\frac{P_{\text{max}}}{N_0B_{\text{s}}}\frac{\chi}{\left(\frac{\epsilon\rho M}{SN}\right)^\frac{\alpha}{2}}\right)}{N}\right)+\Theta\left(\frac{BS}{\epsilon\rho M}\right),P_o= \Theta\left(e^{-\rho}\right).
\end{equation}
Recall that we let 
\begin{equation}
\epsilon\rho=\Theta\left(\left(\frac{S}{M}\right)^{\frac{1}{2-\gamma}}\right).
\end{equation}
This then leads to Theorem 6.

\section{Proof of Theorem 7}

\label{App:ProofThm7}

Considering the same setup as in Appendix \ref{App:ProofThm1} and following the same procedure as in (\ref{eq:g_model_subTC}), the total transport capacity for TX-RX pairs in sub-channel $s$ in time-slot $t$ is given as:
\begin{equation}
\begin{aligned}\label{eq:g_model_subTC_TM7}
&C_{\Gamma_{st}}=\sum_{u\in\Gamma_{st}}r_uc_u=\sum_{u\in\Gamma_{st}}r_u\tau B_{\text{s}}\log_2\left(1+\frac{P_ul_{uu^{(\text{r})}}}{N_0B_{\text{s}}+\sum_{k\in\Gamma_{st},k\neq u}P_kl_{ku^{(\text{r})}}}\right)\\
& \leq r_{w_{st}}\tau B_{\text{s}}\log_2\left(1+\frac{P_{\text{max}}l_{w_{st}w_{st}^{(\text{r})}}}{N_0B_{\text{s}}}\right)+\sum_{u\in\Gamma_{st},u\neq {w_{st}}}r_u\tau B_{\text{s}}\log_2\left(1+\frac{P_ul_{uu^{(\text{r})}}}{N_0B_{\text{s}}+\sum_{k\in\Gamma_{st},k\neq u}P_kl_{ku^{(\text{r})}}}\right),
\end{aligned}
\end{equation}
where $w_{st}$ is the user who has the largest transmission power in $\Gamma_{st}$, i.e., $P_{w_{st}}\geq P_u,\forall u\in\Gamma_{st}$. Then, by denoting the second term in (\ref{eq:g_model_subTC_TM7}) as $C_{\Gamma_{st}}'$ and following the same procedure as in Appendix \ref{App:ProofThm1} with the modification that here $R_0'=\epsilon_0\sqrt{\frac{\alpha_1'q}{SN}}$, we can describe $C_{\Gamma_{st}}'$ as
\begin{equation}
\begin{aligned}\label{eq:g_model_subTC_1_TM7}
C_{\Gamma_{st}}'&=\tau B_{\text{s}}\left[\sum_{u\in\Gamma^{R_0'}_{st}}r_uc_u+\log_2(e)\sum_{j>0}\sum_{u\in\Gamma_{st}^{R_j'}}r_u\log_e\left(1+\frac{P_ul_{uu^{(\text{r})}}}{\sum_{k\in\Gamma_{st},k\neq u}P_kl_{ku^{(\text{r})}}}\right)\right].
\end{aligned}
\end{equation}
Subsequently, by following the procedure from (\ref{eq:g_model_subTC_1}) to
(\ref{eq:g_model_subTC_3}), we can obtain:
\begin{equation}
\begin{aligned}\label{eq:g_model_subTC_10_TM7}
&\sum_{j>0}\sum_{u\in\Gamma_{st}^{R_j'}}r_u\log_e\left(1+\frac{P_ul_{uu^{\text{(r)}}}}{\sum_{k\in\Gamma_{st},k\neq u}P_kl_{ku^{\text{(r)}}}}\right)\leq\sum_{j>0}\left(\left(3\sqrt{2}+1\right)\frac{\alpha}{R_j'}+2(2(\sqrt{2}+1))^{\alpha}\frac{1}{R_j'}\right)\\
&=\left(\alpha\left(3\sqrt{2}+1\right)+2(2(\sqrt{2}+1))^{\alpha}\right)\frac{1}{\epsilon_0}\sqrt{\frac{SN}{\alpha_1' q}}.
\end{aligned}
\end{equation}
By combining (\ref{eq:g_model_subTC_TM7}), (\ref{eq:g_model_subTC_1_TM7}), and (\ref{eq:g_model_subTC_10_TM7}), we obtain
\begin{equation}
\begin{aligned}\label{eq:g_model_subTC_final_10_TM7}
C_{\Gamma_{st}}\leq& r_{w_{st}}\tau B_{\text{s}}\log_2\left(1+\frac{P_{\text{max}}l_{w_{st}w_{st}^{\text{(r)}}}}{N_0B_{\text{s}}}\right)+\tau B_{\text{s}}\sum_{u\in\Gamma^{R_0'}_{st}}r_uc_u\\
&+\tau B_{\text{s}}\frac{\log_2(e)}{\epsilon_0}\sqrt{\frac{SN}{\alpha_1' q}}\left(\alpha\left(3\sqrt{2}+1\right)+2(2(\sqrt{2}+1))^{\alpha}\right)
\end{aligned}
\end{equation}
Finally, by using (\ref{eq:g_model_subTC_final_10_TM7}) and summing contributions in all time-slots and sub-channels, we obtain
\begin{equation}
\begin{aligned}\label{eq:g_model_TC_10_TM7}
&C_{\Gamma}\leq B\overline{C}_{\mathcal{W}}+B\overline{C}_{\Gamma_{R_0}}+B\frac{\log_2(e)}{\epsilon_0}\sqrt{\frac{SN}{\alpha_1' q}}\left(\alpha\left(3\sqrt{2}+1\right)+2(2(\sqrt{2}+1))^{\alpha}\right),
\end{aligned}
\end{equation}
where $\mathcal{W}=\lbrace w_{st},\forall s,t\rbrace$ and
\begin{equation}
\begin{aligned}
\overline{C}_{W}&=\frac{\tau B_{\text{s}}\sum_{t}\sum_{s}r_{w_{st}}\log_2\left(1+\frac{P_{\text{max}}l_{w_{st}w_{st}^{\text{(r)}}}}{N_0B_{\text{s}}}\right)}{BT'};
\overline{C}_{\Gamma_{R_0}}&=\frac{\tau B_{\text{s}}\sum_{t}\sum_{s}\sum_{u\in\Gamma^{R_0}_{st}}r_uc_u}{BT'}.
\end{aligned}
\end{equation}

\section{Proof of Theorem 9}

\label{App:ProofThm9}

By using Theorem 8, we obtain the throughput per user in the first time-slot, given as 
\begin{equation}
\lambda_1=\mathcal{O}\left(\frac{B}{N}\log_2\left(1+\frac{P_{\text{max}}}{N_0B_{\text{s}}}\frac{\chi}{\left(\frac{\alpha_1'q}{SN}\right)^\frac{\alpha}{2}}\right)+\frac{BS}{\alpha_1'q}\right).
\end{equation}
Similarly, the throughput per user in the second time-slot is: 
\begin{equation}\label{eq:out_lambda_2_gg1}
\lambda_2\delta' N\Theta\left(\sqrt{\frac{\epsilon'\alpha_1'q}{SN}}\right)=\mathcal{O}\left(B\log_2\left(1+\frac{P_{\text{max}}}{N_0B_{\text{s}}}\frac{\chi}{\left(\frac{\epsilon'\alpha_1'q}{SN}\right)^\frac{\alpha}{2}}\right)\sqrt{\frac{\epsilon'\alpha_1'q}{SN}}+B\sqrt{\frac{SN}{\epsilon'\alpha_1'q}}\right),
\end{equation}
where $\delta'N$ is the number of users that have $d_u\leq\sqrt{\epsilon'} R_0'$ and $\delta'$ is a function of $\epsilon'$ and $R_0'$. By combining the results in both time-slots, it follows that the overall throughput per user is
\begin{equation}
\begin{aligned}\label{eq:out_lambda_final_gg1}
&T(P_o)=\mathcal{O}\left(\frac{\lambda_1 N+ \lambda_2\delta' N }{2N}\right)\\
&=\mathcal{O}\left(\frac{B}{2}\frac{\log_2\left(1+\frac{P_{\text{max}}}{N_0B_{\text{s}}}\frac{\chi}{\left(\frac{\alpha_1'q}{SN}\right)^\frac{\alpha}{2}}\right)}{N}+\frac{B}{2}\frac{S}{\alpha_1'q}\right)+\mathcal{O}\left(\frac{B}{2}\frac{\log_2\left(1+\frac{P_{\text{max}}}{N_0B_{\text{s}}}\frac{\chi}{\left(\frac{\epsilon'\alpha_1'q}{SN}\right)^\frac{\alpha}{2}}\right)}{N}+\frac{B}{2}\frac{S}{\epsilon'\alpha_1'q}\right),
\end{aligned}
\end{equation}
where the outage lower bound is $P_o=\Theta\left(\frac{1}{(\alpha_1')^{\gamma-1}}\right)$.

By observing (\ref{eq:out_lambda_2_gg1}) and (\ref{eq:out_lambda_final_gg1}), we notice that the throughput enhancement is again dependent on $\epsilon'$ and $\delta'$. We thus follow the same approach in Sec. \ref{Sec:Out_Bound_Gen}.B to characterize the enhanced throughput-outage performance. To do this, similar to the approach of proving Corollary 1, we first need to obtain the maximal probability that users can obtain their desired files from users within the distance of $\sqrt{\epsilon'} R_0'$. This is equivalent to finding the minimal outage probability within a cluster whose side length is $\Theta\left(\sqrt{\epsilon'} R_0'\right)$. Then, by Proposition 3 of \cite{lee2020optimal}, we know that the minimal outage probability in this case is
\begin{equation}
\begin{aligned}\label{eq:Pout_sg_gg1}
p_o^{\text{sec}}=& 1+(\gamma-1)e^{-\gamma\left(\frac{1}{C_1}-1\right)}\cdot\left(\frac{C_1}{C_1+C_2}\right)^{\gamma}\cdot\left(\frac{C_2}{C_1+C_2}\right)^{\gamma\frac{ C_2}{C_1}}\cdot\left(\frac{C_2}{C_1}\right)^{\gamma-1}\\
&\qquad-\left(\left(\frac{C_1}{C_2}\right)^{\gamma-1}-\left(\frac{C_1}{C_1+C_2}\right)^{\gamma-1}\right)\cdot\left(\frac{C_2}{C_1}\right)^{\gamma-1},
\end{aligned}
\end{equation} 
where $g_c'=\Theta\left(\epsilon'(R_0')^2N\right)=\Theta\left(\epsilon'\alpha_1'\frac{q}{S}\right)$, $C_2=\frac{q\gamma}{Sg_c'}$, and $C_1$ is the solution of the equation: $C_1=1+C_2\log\left(1+\frac{C_1}{C_2}\right)$. To proceed the analysis, we provide the following Lemma:

{\em Lemma 6:} Suppose $g_c'=o(q)$, where $g_c'=\Theta\left(\epsilon'\alpha_1'\frac{q}{S}\right)$. It follows that $C_1=\sqrt{2}\left(\frac{\epsilon'\alpha_1'}{\gamma}\right)^{\frac{-1}{2}}$, $C_2=\left(\frac{\epsilon'\alpha_1'}{\gamma}\right)^{-1}$, and $\frac{C_1}{C_2}=\sqrt{2}\left(\frac{\epsilon'\alpha_1'}{\gamma}\right)^{\frac{1}{2}}$.
\begin{proof}
See Appendix \ref{App:ProofLemm4}.
\end{proof}

Observe that we have $\epsilon'\alpha_1'=o(1)\to 0$, and therefore $C_1\to\infty$, $C_2\to\infty$, and $\frac{C_1}{C_2}\to 0$. Then by using Lemma 6, we can simplify (\ref{eq:Pout_sg_gg1}) as follows:
\begin{equation}
\begin{aligned}\label{eq:Pout_sg_gg1_2}
&p_o^{\text{sec}}= 1+(\gamma-1)e^{-\gamma\left(\frac{1}{C_1}-1\right)}\cdot\left(\frac{C_1}{C_1+C_2}\right)^{\gamma}\cdot\left(\frac{C_2}{C_1+C_2}\right)^{\gamma\frac{ C_2}{C_1}}\cdot\left(\frac{C_2}{C_1}\right)^{\gamma-1}\\
&\qquad-\left(\left(\frac{C_1}{C_2}\right)^{\gamma-1}-\left(\frac{C_1}{C_1+C_2}\right)^{\gamma-1}\right)\cdot\left(\frac{C_2}{C_1}\right)^{\gamma-1}\\
&=1+(\gamma-1)e^{\left(\frac{-\gamma}{C_1}\right)}e^{\gamma}\left(\frac{C_1}{C_1+C_2}\right)^{\gamma}\underbrace{\left(1-\frac{C_1}{C_1+C_2}\right)^{\gamma\frac{ C_2}{C_1}}}_{\stackrel{(a)}{=}e^{-\gamma}}\left(\frac{C_2}{C_1}\right)^{\gamma-1}-\left(1-\left(\frac{C_2}{C_1+C_2}\right)^{\gamma-1}\right)\\
&=1+(\gamma-1)e^{\left(\frac{-\gamma}{C_1}\right)}\left(\frac{C_1}{C_1+C_2}\right)^{\gamma}\left(\frac{C_2}{C_1}\right)^{\gamma-1}-\left(1-\left(1-\frac{C_1}{C_1+C_2}\right)^{\gamma-1}\right)\\
&\stackrel{(b)}{=}1+(\gamma-1)\left(1-\frac{\gamma}{C_1}\right)\left(\frac{C_1}{C_1+C_2}\right)^{\gamma}\left(\frac{C_2}{C_1}\right)^{\gamma-1}-\left(1-\left(1-(\gamma-1)\frac{C_1}{C_1+C_2}\right)\right)-\Theta\left(\left(\frac{C_1}{C_2}\right)^2\right)\\
&=1+(\gamma-1)\left(\frac{C_1}{C_1+C_2}\right)^{\gamma}\left(\frac{C_2}{C_1}\right)^{\gamma-1}-\frac{\gamma(\gamma-1)}{C_1}\left(\frac{C_1}{C_1+C_2}\right)^{\gamma}\left(\frac{C_2}{C_1}\right)^{\gamma-1}-(\gamma-1)\frac{C_1}{C_1+C_2}-\Theta\left(\left(\frac{C_1}{C_2}\right)^2\right)\\
&=1+(\gamma-1)\left(\frac{C_2}{C_1+C_2}\right)^{\gamma}\left(\frac{C_1}{C_2}\right)-\frac{\gamma(\gamma-1)}{C_2}\left(\frac{C_2}{C_1+C_2}\right)^{\gamma}\left(\frac{C_1}{C_2}\right)-(\gamma-1)\frac{C_1}{C_1+C_2}-\Theta\left(\left(\frac{C_1}{C_2}\right)^2\right)\\
&=1+(\gamma-1)\left(1-\frac{C_1}{C_1+C_2}\right)^{\gamma}\left(\frac{C_1}{C_2}\right)\\
&\quad-\frac{\gamma(\gamma-1)}{C_2}\left(1-\frac{C_1}{C_1+C_2}\right)^{\gamma}\left(\frac{C_1}{C_2}\right)-(\gamma-1)\frac{C_1}{C_1+C_2}-\Theta\left(\left(\frac{C_1}{C_2}\right)^2\right)\\
&\stackrel{(c)}{=}1+(\gamma-1)\left(1-\frac{\gamma C_1}{C_1+C_2}\right)\left(\frac{C_1}{C_2}\right)-\frac{\gamma(\gamma-1)}{C_2}\left(1-\frac{\gamma C_1}{C_1+C_2}\right)\left(\frac{C_1}{C_2}\right)-(\gamma-1)\frac{C_1}{C_1+C_2}-\Theta\left(\left(\frac{C_1}{C_2}\right)^2\right)\\
&=1+(\gamma-1)\frac{C_1}{C_2}-\frac{\gamma C_1}{C_1+C_2}\frac{C_1}{C_2}-\frac{\gamma(\gamma-1)}{C_2}\frac{C_1}{C_2}+\frac{\gamma(\gamma-1)}{C_2}\frac{\gamma C_1}{C_1+C_2}\frac{C_1}{C_2}-(\gamma-1)\frac{C_1}{C_1+C_2}-\Theta\left(\left(\frac{C_1}{C_2}\right)^2\right)\\
&\stackrel{(d)}{=}1+(\gamma-1)\frac{C_1}{C_2}-(\gamma-1)\frac{C_1}{C_1+C_2}-\Theta\left(\left(\frac{C_1}{C_2}\right)^2\right)=1+(\gamma-1)\frac{C_1^2}{C_2(C_1+C_2)}-\Theta\left(\left(\frac{C_1}{C_2}\right)^2\right)\\
&\stackrel{(e)}{=}1-\Theta\left(\left(\frac{C_1}{C_2}\right)^2\right)=1-\Theta\left(\frac{\epsilon'\alpha_1'}{\gamma}\right),
\end{aligned}
\end{equation} 
where $(a)$ is because $\frac{C_2}{C_1}\to\infty$; $(b)$ is because $\frac{C_1}{C_1+C_2}\to 0$ and $\frac{-\gamma}{C_1}\to 0$; $(c)$ is again because $\frac{C_1}{C_1+C_2}\to 0$; $(d)$ and $(e)$ are because $C_1,C_2\to\infty$, $\frac{C_1}{C_2}\to 0$, $C_1=o(C_2)$, and $p_o\leq 1$. From (\ref{eq:Pout_sg_gg1_2}), we can then obtain the probability of having a TX-RX pair to be within $\sqrt{\epsilon'}R_0'$ is
\begin{equation}\label{eq:Phit_2_gg1}
p_{\text{h}}^{\text{sec}}=\Theta\left(\frac{\epsilon'\alpha_1'}{\gamma}\right),
\end{equation}
where $\epsilon'\alpha'=o(1)$. This leads to 
\begin{equation}\label{eq:delta_bound_gg1}
\delta'=\Theta\left(\epsilon'\alpha_1'\right).
\end{equation}
By using (\ref{eq:out_lambda_2_gg1}), (\ref{eq:out_lambda_final_gg1}), and (\ref{eq:delta_bound_gg1}), we complete the proof.

\section{Proof of Lemma 6}

\label{App:ProofLemm4}

Since $g_c'=\epsilon'\alpha_1'\frac{q}{S}$ and $C_2=\frac{q\gamma}{Sg_c'}$, we obtain $C_2=\left(\frac{\epsilon'\alpha_1'}{\gamma}\right)^{-1}$. Then, observe that $C_1=1+C_2\log\left(1+\frac{C_1}{C_2}\right)$. Hence, 
\begin{equation}
\frac{C_1}{C_2}=\frac{1}{C_2}+\log\left(1+\frac{C_1}{C_2}\right).
\end{equation}
Let $x=\frac{C_1}{C_2}$. The above equation becomes:
\begin{equation}
x=\left(\frac{\epsilon'\alpha_1'}{\gamma}\right)+\log(1+x).
\end{equation}
Assume $x\to 0$. Then, by Taylor's expansion, we can have the approximation: $\log(1+x)\approx x-\frac{x^2}{2}$. It follows that we obtain
\begin{equation}
\frac{x^2}{2}=\left(\frac{\epsilon'\alpha_1'}{\gamma}\right).
\end{equation}
Therefore, $x=\left(\frac{2\epsilon'\alpha_1'}{\gamma}\right)^{\frac{1}{2}}$.
Recall that $x=\frac{C_1}{C_2}$. This leads to $C_1=\sqrt{2}\left(\frac{\epsilon'\alpha_1'}{\gamma}\right)^{\frac{-1}{2}}$. Note that since $g_c'=o(q)$ and $g_c'=\Theta\left(\epsilon'\alpha_1'\frac{q}{S}\right)$, we know $x=\left(\frac{2\epsilon'\alpha_1'}{\gamma}\right)^{\frac{1}{2}}\to 0$. It follows that our assumption for the adopted approximation is valid. This concludes the proof.

\section{Proof of Theorem 11}

\label{App:ProofThm11}

Recall that the number of users $N_{\text{cluster}}$ in a cluster follows the Poisson distribution whose mean value is $\frac{\alpha_1q}{S}$. Then, according to Corollary 3 of \cite{lee2020optimal}, when $\alpha_1$ is sufficiently large, the outage probability is
\begin{equation}
p_o=\Theta\left(\frac{1}{(\alpha_1)^{\gamma-1}}\right),
\end{equation}
indicating that the outage probability can be negligibly small or converging to zero. Furthermore, since the mean number of users is $\frac{\alpha_1q}{S}\to \infty$ and the outage probability is either negligibly small or converging to zero, we can conclude that the probability that no user can obtain their desired files from users in the same cluster converges to zero. It follows by repeating the same procedure as in Appendix \ref{App:ProofThm5}, we shall obtain the achievable throughput per user generated at the first timeslot as:
\begin{equation}
T_1=\frac{B}{2N}\log_2\left(1+\frac{P_{\text{max}}}{N_0B}\frac{\chi}{\left(\frac{\alpha_1 q}{SN}\right)^\frac{\alpha}{2}}\right);
\end{equation}
and the achievable throughput per user generated at the second timeslot as:
\begin{equation}
T_2\geq\frac{1}{N}\left\lfloor\frac{SN}{\alpha_1 q}\right\rfloor\cdot\frac{B}{(2(K+1))^2}\log_2\left(1+\vartheta\right)=\Theta\left(\frac{BS}{\alpha_1 q}\right).
\end{equation}
By combining the throughput generated at the first and second time-slots, we obtain the following achievable throughput per user:
\begin{equation}
T(P_o)=\underbrace{\Theta\left(\frac{1}{N}\log_2\left(1+\frac{P_{\text{max}}}{N_0B}\frac{\chi}{\left(\frac{\alpha_1 q}{SN}\right)^\frac{\alpha}{2}}\right)\right)}_{\text{come from the first timeslot}}+\underbrace{\Theta\left(\frac{BS}{\alpha_1 q}\right)}_{\text{come from the second timeslot}},
\end{equation}
where the corresponding achievable outage probability is 
\begin{equation}
P_o= \Theta\left(\frac{1}{(\alpha_1)^{\gamma-1}}\right).
\end{equation}

\section{Proof of Theorem 12}

\label{App:ProofThm12}

For the first caching subspace, we consider the caching policy proposed in Theorem 1 of \cite{lee2020optimal} with $g_{c,1}(M)=\frac{2\alpha_1 q}{S}$. Then, for the second caching subspace, we adopt the caching policy proposed in Theorem 1 of \cite{lee2020optimal} and let $g_{c,2}(M)=\frac{2\epsilon\alpha_1 q}{S}$. Therefore, by using Theorem 11 with $d_1=\sqrt{\frac{\alpha_1 q}{SN}}$, we obtain the achievable throughput-outage performance for first time-slot as:
\begin{equation}
\begin{aligned}
&T_1(P_o)= \Theta\left(\frac{1}{N}\log_2\left(1+\frac{P_{\text{max}}}{N_0B}\frac{\chi}{\left(\frac{\alpha_1 q}{SN}\right)^\frac{\alpha}{2}}\right)+\frac{S}{\alpha_1 q}\right),P_o= \Theta\left(\frac{1}{(\alpha_1)^{\gamma-1}}\right),
\end{aligned}
\end{equation}
where $P_o$ can be negligibly small or converging to zero. Note that since we let $d_2=\sqrt{\frac{2\epsilon\alpha_1 q}{SN}}$ in the second timeslot, the probability that a user can find their desired in the cluster is lower bounded by the cache-hit probability when considering the second caching subspace. Then, according to Proposition 3 of \cite{lee2020optimal} and derivations in Appendix \ref{App:ProofThm9}, the cache-hit probability $p_{\text{h}}^{\text{sec}}$ is given by (\ref{eq:Phit_2_gg1}), where
\begin{equation}
p_{\text{h}}^{\text{sec}}=\Theta\left(\frac{\epsilon\alpha_1}{\gamma}\right).
\end{equation}
Then, by letting 
\begin{equation}\label{eq:Ach_re_gg1}
\epsilon\alpha_1=\Theta\left(\left(\frac{S}{q}\right)^{\frac{1}{2}}\right),
\end{equation}
the average number of users that can find the desired files in a cluster with side length $d_2=\sqrt{\frac{2\epsilon\alpha_1 q}{SN}}$ is given as
\begin{equation}
\Theta\left(\frac{2\epsilon\alpha_1 q}{S}p_{\text{h}}^{\text{sec}}\right)=\Theta\left(1\right),
\end{equation}
which can be very large if we let (\ref{eq:Ach_re_gg1}) to have a large constant factor, namely, $\epsilon\alpha_1=C_{\text{sec}}\left(\frac{S}{q}\right)^{\frac{1}{2}}$, where $C_{\text{sec}}$ is a large constant. Hence, the probability of activating a cluster with side length $d_2=\sqrt{\frac{2\epsilon\alpha_1 q}{SN}}$ can be close to 1. Then, since we have 
\begin{equation}
\frac{1}{d_2^2}=\frac{N S}{2\alpha_1\epsilon q}=\frac{N}{2}\left(\frac{S}{q}\right)^{\frac{1}{2}}
\end{equation}
number of clusters, by Proposition 1, the total throughput in the second time-slot is 
\begin{equation}
\Theta\left(\frac{BN}{2}\left(\frac{S}{q}\right)^{\frac{1}{2}}\right).
\end{equation}
Note that the number of users that can receive files in the second time-slot is
\begin{equation}
Np_{\text{h}}^{\text{sec}}=\Theta\left(N\left(\frac{S}{q}\right)^{\frac{1}{2}}\right).
\end{equation}
It follows that the throughput per user for users that can receive files in the second time-slot is
\begin{equation}
T_2=\Theta\left(\frac{B}{2}\frac{\log_2\left(1+\frac{P_{\text{max}}}{N_0B_{\text{s}}}\frac{\chi}{\left(\left(\frac{q}{S}\right)^{\frac{1}{2}}\frac{1}{N}\right)^\frac{\alpha}{2}}\right)}{\left(\left(\frac{S}{q}\right)^{\frac{1}{2}}\right)N}\right)+\Theta(B).
\end{equation}
Finally, by using 
\begin{equation}
T=\Theta\left(\frac{T_1 N+ T_2Np_{\text{h}}^{\text{sec}} }{2N}\right)
\end{equation} 
and above discussions, we complete the proof.

\bibliographystyle{IEEEtran}
\bibliography{Bib_D2D_IT_[2020_09_30]}

\begin{thebibliography}{10}
\providecommand{\url}[1]{#1}
\csname url@samestyle\endcsname
\providecommand{\newblock}{\relax}
\providecommand{\bibinfo}[2]{#2}
\providecommand{\BIBentrySTDinterwordspacing}{\spaceskip=0pt\relax}
\providecommand{\BIBentryALTinterwordstretchfactor}{4}
\providecommand{\BIBentryALTinterwordspacing}{\spaceskip=\fontdimen2\font plus
\BIBentryALTinterwordstretchfactor\fontdimen3\font minus
  \fontdimen4\font\relax}
\providecommand{\BIBforeignlanguage}[2]{{%
\expandafter\ifx\csname l@#1\endcsname\relax
\typeout{** WARNING: IEEEtran.bst: No hyphenation pattern has been}%
\typeout{** loaded for the language `#1'. Using the pattern for}%
\typeout{** the default language instead.}%
\else
\language=\csname l@#1\endcsname
\fi
#2}}
\providecommand{\BIBdecl}{\relax}
\BIBdecl

\bibitem{lee2021throughput}
M.-C. Lee, A.~F. Molisch, and M.~Ji, ``Throughput--outage scaling laws for
  wireless single-hop {D2D} caching networks with physical models,'' in
  \emph{2021 IEEE International Conference on Communications (ICC)}.\hskip 1em
  plus 0.5em minus 0.4em\relax IEEE, 2021, (to appear).

\bibitem{Cisco_2017}
``Cisco virtual networking index: Global mobile data traffic forecast update,
  2017-2022,'' San Jose, CA, USA, Tech. Rep.

\bibitem{wang2017survey}
S.~Wang, X.~Zhang, Y.~Zhang, L.~Wang, J.~Yang, and W.~Wang, ``A survey on
  mobile edge networks: Convergence of computing, caching and communications,''
  \emph{IEEE Access}, vol.~5, pp. 6757--6779, 2017.

\bibitem{zhang20196g}
Z.~Zhang, Y.~Xiao, Z.~Ma, M.~Xiao, Z.~Ding, X.~Lei, G.~K. Karagiannidis, and
  P.~Fan, ``6g wireless networks: Vision, requirements, architecture, and key
  technologies,'' \emph{IEEE Veh. Technol. Mag.}, vol.~14, no.~3, pp. 28--41,
  2019.

\bibitem{Andrews:5G}
J.~G. Andrews, S.~Buzzi, W.~Choi, and et. al., ``What will 5g be?'' \emph{IEEE
  J. Sel. Areas Commun.}, vol.~32, no.~6, pp. 1065--1082, January 2014.

\bibitem{Gol:femtocaching}
N.~Golrezaei, A.~F. Molisch, A.~G. Dimakis, and G.~Caire, ``Femtocaching and
  device-to-device collaboration: A new architecture for wireless video
  distribution,'' \emph{IEEE Commmun. Mag.}, vol.~51, no.~4, pp. 142--149,
  April 2013.

\bibitem{Ji:Dcache}
M.~Ji, G.~Caire, and A.~F. Molisch, ``Wireless device-to-device caching
  networks: Basic principles and system performance,'' \emph{IEEE J. Sel. Area
  Commun.}, vol.~34, no.~1, pp. 176--189, January 2016.

\bibitem{li2018survey}
L.~Li, G.~Zhao, and R.~S. Blum, ``A survey of caching techniques in cellular
  networks: Research issues and challenges in content placement and delivery
  strategies,'' \emph{IEEE Communications Surveys \& Tutorials}, vol.~20,
  no.~3, pp. 1710--1732, 2018.

\bibitem{Maddah-Ali:CCache}
M.~A. Maddah-Ali and U.~Niessen, ``Fundamental limits of caching,'' \emph{IEEE
  Trans. Inf. Theory}, vol.~60, no.~5, pp. 2856--2867, May 2014.

\bibitem{Ji:Th_Out_toff}
M.~Ji, G.~Gaire, and A.~F. Molisch, ``The throughput-outage tradeoff of
  wireless one-hop caching networks,'' \emph{IEEE Trans. Inf. Theory}, vol.~61,
  no.~12, pp. 6833--6859, December 2015.

\bibitem{lee2019throughput}
M.-C. Lee, M.~Ji, A.~F. Molisch, and N.~Sastry, ``Throughput-outage analysis
  and evaluation of cache-aided {D2D} networks with measured popularity
  distributions,'' \emph{IEEE Trans. on Wireless Commun.}, vol.~18, no.~11, pp.
  5316--5332, November 2019.

\bibitem{lee2020optimal}
M.-C. Lee, M.~Ji, and A.~F. Molisch, ``Optimal throughput--outage analysis of
  cache-aided wireless multi-hop {D2D} networks,'' \emph{IEEE Trans. Commun.},
  2020, (In press).

\bibitem{ahmed2019video}
I.~Ahmed, M.~H. Ismail, and M.~S. Hassan, ``Video transmission using
  device-to-device communications: A survey,'' \emph{IEEE Access}, vol.~7, pp.
  131\,019--131\,038, 2019.

\bibitem{mehrabi2019device}
M.~Mehrabi, D.~You, V.~Latzko, H.~Salah, M.~Reisslein, and F.~H.~P. Fitzek,
  ``Device-enhanced mec: Multi-access edge computing (mec) aided by end device
  computation and caching: A survey,'' \emph{IEEE Access}, vol.~7, pp.
  166\,079--166\,108, 2019.

\bibitem{prerna2020device}
D.~Prerna, R.~Tekchandani, and N.~Kumar, ``Device-to-device content caching
  techniques in 5g: A taxonomy, solutions, and challenges,'' \emph{Computer
  Communications}, 2020.

\bibitem{li2017collaborative}
X.~Li, X.~Wang, K.~Li, Z.~Han, and V.~C.~M. Leung, ``Collaborative multi-tier
  caching in heterogeneous networks: Modeling, analysis, and design,''
  \emph{IEEE Trans. Wireless Commun.}, vol.~16, no.~10, pp. 6926--6939, 2017.

\bibitem{hu2018mobility}
R.~Q. Hu \emph{et~al.}, ``Mobility-aware edge caching and computing in vehicle
  networks: A deep reinforcement learning,'' \emph{IEEE Trans. on Veh.
  Technol.}, vol.~67, no.~11, pp. 10\,190--10\,203, 2018.

\bibitem{gupta2000capacity}
P.~Gupta and P.~R. Kumar, ``The capacity of wireless networks,'' \emph{IEEE
  Trans. on inf. theory}, vol.~46, no.~2, pp. 388--404, 2000.

\bibitem{agarwal2004capacity}
A.~Agarwal and P.~R. Kumar, ``Capacity bounds for ad hoc and hybrid wireless
  networks,'' \emph{ACM SIGCOMM Computer Communication Review}, vol.~34, no.~3,
  pp. 71--81, 2004.

\bibitem{xue2006scaling}
F.~Xue and P.~R. Kumar, ``Scaling laws for ad hoc wireless networks: an
  information theoretic approach,'' \emph{Foundations and
  Trends{\textregistered} in Networking}, vol.~1, no.~2, pp. 145--270, 2006.

\bibitem{franceschetti2007closing}
M.~Franceschetti, O.~Dousse, D.~N. Tse, and P.~Thiran, ``Closing the gap in the
  capacity of wireless networks via percolation theory,'' \emph{IEEE Trans.
  Inf. Theory}, vol.~53, no.~3, pp. 1009--1018, 2007.

\bibitem{el2006optimal_I}
A.~El~Gamal, J.~Mammen, B.~Prabhakar, and D.~Shah, ``Optimal throughput-delay
  scaling in wireless networks-part i: The fluid model,'' \emph{IEEE Trans. on
  Inf. Theory}, vol.~52, no.~6, pp. 2568--2592, 2006.

\bibitem{shakkottai2010multicast}
S.~Shakkottai, X.~Liu, and R.~Srikant, ``The multicast capacity of large
  multihop wireless networks,'' \emph{IEEE/ACM Trans Networking}, vol.~18,
  no.~6, pp. 1691--1700, 2010.

\bibitem{ozgur2007hierarchical}
A.~Ozgur, O.~L{\'e}v{\^e}que, and D.~N. Tse, ``Hierarchical cooperation
  achieves optimal capacity scaling in ad hoc networks,'' \emph{IEEE Trans.
  inf. theory}, vol.~53, no.~10, pp. 3549--3572, 2007.

\bibitem{hara2001effective}
T.~Hara, ``Effective replica allocation in ad hoc networks for improving data
  accessibility,'' in \emph{IEEE INFOCOM 2001}, vol.~3.\hskip 1em plus 0.5em
  minus 0.4em\relax IEEE, 2001, pp. 1568--1576.

\bibitem{cohen2002replication}
E.~Cohen and S.~Shenker, ``Replication strategies in unstructured peer-to-peer
  networks,'' \emph{ACM SIGCOMM Computer Communication Review}, vol.~32, no.~4,
  pp. 177--190, 2002.

\bibitem{hara2006data}
T.~Hara and S.~K. Madria, ``Data replication for improving data accessibility
  in ad hoc networks,'' \emph{IEEE transactions on mobile computing}, vol.~5,
  no.~11, pp. 1515--1532, 2006.

\bibitem{zhao2010cooperative}
J.~Zhao, P.~Zhang, G.~Cao, and C.~R. Das, ``Cooperative caching in wireless p2p
  networks: Design, implementation, and evaluation,'' \emph{IEEE Trans.
  Parallel and Distributed Systems}, vol.~21, no.~2, pp. 229--241, 2010.

\bibitem{Golrezaei:Cach_scale}
N.~Golrezaei, A.~D. Dimakis, and A.~F. Molisch, ``Scaling behavior for
  device-to-device communications with distributed caching,'' \emph{IEEE Trans.
  Inf. Theory}, vol.~60, no.~7, pp. 4286--4298, July 2014.

\bibitem{ji2013fundamental}
M.~Ji, G.~Caire, and A.~F. Molisch, ``Fundamental limits of distributed caching
  in d2d wireless networks,'' in \emph{2013 IEEE Information Theory Workshop
  (ITW)}.\hskip 1em plus 0.5em minus 0.4em\relax IEEE, 2013, pp. 1--5.

\bibitem{gitzenis2012asymptotic}
S.~Gitzenis, G.~S. Paschos, and L.~Tassiulas, ``Asymptotic laws for joint
  content replication and delivery in wireless networks,'' \emph{IEEE Trans.
  Inf. Theory}, vol.~59, no.~5, pp. 2760--2776, 2012.

\bibitem{jeon2017wireless}
S.-W. Jeon, S.-N. Hong, M.~Ji, G.~Caire, and A.~F. Molisch, ``Wireless multihop
  device-to-device caching networks,'' \emph{IEEE Trans. Inf. Theory}, vol.~63,
  no.~3, pp. 1662--1676, 2017.

\bibitem{qiu2019popularity}
L.~Qiu and G.~Cao, ``Popularity-aware caching increases the capacity of
  wireless networks,'' \emph{IEEE Trans. Mobile Comput.}, vol.~19, no.~1, pp.
  173--187, 2019.

\bibitem{guo2017achievable}
J.~Guo, J.~Yuan, and J.~Zhang, ``An achievable throughput scaling law of
  wireless device-to-device caching networks with distributed mimo and
  hierarchical cooperations,'' \emph{IEEE Trans. on Wireless Commun.}, vol.~17,
  no.~1, pp. 492--505, 2017.

\bibitem{ren2021scaling}
J.~Ren, D.~Li, L.~Zhang, and G.~Zhang, ``Scaling performance analysis and
  optimization based on the node spatial distribution in mobile content-centric
  networks,'' \emph{Wireless Communications and Mobile Computing}, vol. 2021,
  Jan. 2021.

\bibitem{ji2017fundamental}
M.~Ji, R.-R. Chen, G.~Caire, and A.~F. Molisch, ``Fundamental limits of
  distributed caching in multihop d2d wireless networks,'' in \emph{2017 IEEE
  International Symposium on Information Theory (ISIT)}, 2017, pp. 2950--2954.

\bibitem{Ji:Fund_D2D}
M.~Ji, G.~Caire, and A.~F. Molisch, ``Fundamental limits of caching in wireless
  d2d networks,'' \emph{IEEE Trans. Inf. Theory}, vol.~62, no.~2, pp. 849--869,
  February 2016.

\bibitem{naderializadeh2017fundamental}
N.~Naderializadeh, M.~A. Maddah-Ali, and A.~S. Avestimehr, ``Fundamental limits
  of cache-aided interference management,'' \emph{IEEE Transactions on
  Information Theory}, vol.~63, no.~5, pp. 3092--3107, 2017.

\bibitem{yapar2019optimality}
{\c{C}}.~Yapar, K.~Wan, R.~F. Schaefer, and G.~Caire, ``On the optimality of
  d2d coded caching with uncoded cache placement and one-shot delivery,''
  \emph{IEEE Trans. Commun.}, vol.~67, no.~12, pp. 8179--8192, 2019.

\bibitem{zhang2021cache}
X.~Zhang, N.~Woolsey, and M.~Ji, ``Cache-aided interference management using
  hypercube combinatorial design with reduced subpacketizations and order
  optimal sum-degrees of freedom,'' \emph{IEEE Trans. Wireless Commun.}, 2021.

\bibitem{Zhang:D2D_Schedule}
L.~Zhang, M.~Xiao, G.~Wu, and S.~Li, ``Efficient scheduling and power
  allocation for d2d-assisted wireless caching networks,'' \emph{IEEE Trans.
  Commun.}, vol.~64, no.~6, pp. 2438--2452, June 2016.

\bibitem{chen2017optimal}
B.~Chen, C.~Yang, and Z.~Xiong, ``Optimal caching and scheduling for
  cache-enabled d2d communications,'' \emph{IEEE Commun. Lett.}, vol.~21,
  no.~5, pp. 1155--1158, 2017.

\bibitem{lee2018cachingTWC}
M.-C. Lee and A.~F. Molisch, ``Caching policy and cooperation distance design
  for base station assisted wireless d2d caching networks: Throughput and
  energy efficiency optimization and trade-off,'' \emph{IEEE Trans. on Wireless
  Commun.}, vol.~17, no.~11, pp. 7500--7514, November 2018.

\bibitem{choi2020joint}
M.~Choi, A.~F. Molisch, and J.~Kim, ``Joint distributed link scheduling and
  power allocation for content delivery in wireless caching networks,''
  \emph{IEEE Trans. on Wireless Commun.}, vol.~19, no.~12, pp. 7810--7824,
  2020.

\bibitem{Blaszczyszyn:fcache}
B.~Blaszczyszyn and A.~Giovanidis, ``Optimal geographic caching in cellular
  networks,'' June 2015.

\end{thebibliography}
%

\end{document}